\documentclass[11pt,draftcls,onecolumn]{IEEEtran}

\IEEEoverridecommandlockouts

\usepackage[doublespacing]{setspace}
\usepackage{tocloft}
\usepackage[margin=1in]{geometry}
\usepackage[cmex10]{amsmath}
\usepackage{textcomp}
\usepackage{amssymb}

\usepackage{cite}
\usepackage{graphicx}
\usepackage{color}

\usepackage{mathtools}

\newtheorem{theorem}{Theorem}
\newtheorem{lemma}{Lemma}

\newcommand{\Xc}{\mathcal{X}}

\begin{document}
\title{Achievable Rates for Binary Two-hop Channel with Energy Harvesting Relay With Finite Battery} %
\author{\normalsize Ali H. Abdollahi Bafghi, Mahtab Mirmohseni, and Mohammad Reza Aref\\

\thanks{This work was supported in part by the Iran National Science Foundation under Grant 92-32575. This work has been presented in part in 5th Iran Workshop on Communication and Information Theory 2017 \cite{Abdollahi}.}
\thanks{Authors are with Information Systems and Security Lab (ISSL), Department of Electrical Engineering, Sharif University of Technology, Tehran, Iran (email: {aliabdolahi}@ee.sharif.edu,\{mirmohseni,aref\}@sharif.edu)}

}


\maketitle

\begin{abstract}
We study the problem of joint information and energy transfer in a binary two-hop channel with an energy harvesting relay. We consider a finite battery size at the relay and energy loss in transmitting energy. In other words, to be able to send an energy-contained symbol, the relay must receive multiple energy-contained symbols. Thus, we face a kind of channel with memory. We model the energy saved in the battery as the channel state with the challenge that the receiver does not know the channel state. We propose two different achievable schemes, the first one is based on state-dependent superposition coding and the second one is based on the equivalent timing channel approach. Both of our schemes are based on block Markov coding and backward decoding techniques. Due to these two approaches, we find achievable rates with a single-letter expression for the model.
\end{abstract}


\section{Introduction}
Nowadays, energy consumption becomes an important design factor in communication systems because of financial reasons and environmental concerns. There would be a large number of devices and thus large energy consumption in next-generation (6G) networks, thus, energy efficiency will play an important role in these networks \cite{Buzzi}.
One alternative technique used for energy management is energy harvesting. Energy harvesting enables wireless networks to exploit environmental energies to increase energy efficiency.
This promising method is introduced in two main directions: (i) in the first direction, energy is harvested from environmental sources like wind and sunlight. The main characteristic of this setup is the sporadic nature of the harvested energy, which makes the exact design and analysis rather difficult; (ii) in the second direction, known as radio frequency (RF) energy harvesting, energy is harvested from the electromagnetic waves in the space. This technique is promising with an increasing demand profile and some commercialized products \cite{lu2015wireless}.
Due to less randomness in obtaining the harvested energy from RF signals, the analysis and strategies could be simpler compared with the harvesting from the environment.

The RF energy can be transferred concurrently with the information signal in a wireless system, proposed as simultaneous wireless information and power transfer (SWIPT) \cite{lu2015wireless,Buzzi}. A convenient scenario, in this case, is to save the received energy to be used for future transmissions. The optimal design of the encoder and decoder is a challenging problem because the memory appears in the system. Also, the existing works show an inherent trade-off in transmitting energy and information simultaneously \cite{lu2015wireless}. An information-theoretic model helps to understand the interplay between energy and information transmission and also to obtain the optimal coding structure in this scenario.
Because of the complexity of models (especially channels with memory), the fundamental limits of these systems are not noticed widely. In particular, it is not clear what are the properties of an optimal code in these systems (the systems that transmit information and energy simultaneously).

From an information-theoretic perspective, the energy harvesting systems have been studied in \cite{Ozel}-\cite{Hassibi}.
The capacity of a point to point additive white Gaussian noise (AWGN) channel with {random} power available at a transmitter with infinite battery storage was studied by Ozel and Ulukus \cite{Ozel}. They showed that the capacity in this channel is equal to the capacity of an AWGN channel with average power constraint equal to average recharge rate. A bounded-gap approximation of the capacity of an energy harvesting AWGN channel at the presence of an exogenous stochastic energy harvesting process was derived by Shaviv et. al. in \cite{Nguyen}.
The capacity of an energy harvesting channel, where a binary energy process is available at the transmitter and battery size is one, was derived by Tutuncuoglu et. al. in \cite{ulukus}. They used an equivalent timing channel to convert the problem to a memoryless channel with additive noise.
Mao and Hassibi, using techniques of channels with side information and finite-state channels, derived inner and outer bounds on the capacity of discrete energy harvesting channels \cite{Hassibi}.

The fundamental limits of SWIPT are studied in some pioneering works \cite{Gastpar}-\cite{Fouladgar}.
The capacity of a channel with a constraint on output was determined by Gastpar, which can be adopted to a channel with energy reception constraint \cite{Gastpar}. These results were also extended to Gaussian relay and multiple access channels \cite{Gastpar}. The capacity-energy function was introduced by Varshney to study the problem of transmission of optimal rate with a constraint on minimum received energy \cite{Varshney}. One of the encoding methods for simultaneous information and energy transfer is run-length limited (RLL) codes, which were analyzed by Fouladgar et. al. \cite{Fouladgar}.

In the works mentioned above, the interaction between energy and information was not studied when the harvesting node has to use the received information and energy for future transmissions.
Interactive simultaneous transfer of information and energy was studied by Popovski et. al. in \cite{popovski}, where an orthogonal two-way channel is considered, in which each node has a finite battery and transmits its message with harvested energy from the received signal. A coding scheme was proposed and inner and outer bounds were derived for the capacity region in \cite{popovski}. The capacity-energy function was introduced for a multi-hop channel with an energy harvesting relay by Fouladgar and Simeone \cite{Simeone}, where infinite storage for harvested energy at the relay was considered. They also derived the capacity of a multiple access channel with a constraint on the received power \cite{Simeone}. Now the main question is that how a finite battery at the relay may change the fundamental limits of an energy harvesting relay channel.

In this paper, we consider a two-hop channel with an energy harvesting relay, where the transmitter jointly transfers information and energy to the relay. The harvested energy at the relay is used to re-transmit the data to the receiver. We assume finite battery size at the relay. In addition, we consider energy loss in transmitting energy in our model. In fact, the relay must receive multiple energy-contained symbols to be able to send one energy-contained symbol. These limitations at the relay turn the problem to the transmission over a channel \emph{with states}, where the state shows the energy level at the relay's battery.
Hence, we face a kind of channel with memory.
Thus, the main questions are which rates would be achievable in these models and which coding structures will achieve those rates. One of the main challenges in the coding design for this model is to make the receiver be able to decode the message without knowing the sequence of states.
We model the energy stored in the relay's battery as a channel state with the challenge that the receiver does not know the channel state. In the following, we summarize the main results of this paper:

\begin{itemize}

\item
In the first step, we assume that the energy loss is deterministic. We study the problem for three cases, i) there is a memoryless noise in the second hop only, ii) there is a memoryless noise in the first hop only, and iii) there is memoryless noise in both hops. For the first and third cases, we propose a state-dependent coding scheme based on block Markov coding at the transmitter and backward decoding at the receiver and derive a single letter achievable rate for it.

\item
For the second case, we model the channel between the relay and the receiver (second hop) by an equivalent timing channel and propose a block Markov coding and backward decoding based achievability scheme. We derived a single letter achievable rate for this scenario.
\item
At last, we extend the results of the third case to random energy loss using the first coding scheme and derive a single letter achievable rate for it.

\end{itemize}
The remainder of this paper is organized as follows.
In Section \ref{section2}, we introduce the system model of our problem. In Section \ref{section3}, we present our coding scheme for two-hop and energy harvesting relay channel with finite battery with noise in the second hop. In Section \ref{section4}, we present a new coding scheme for the case when noise is only in the first hop. In Section \ref{section5}, we extend the results of section \ref{section3} for the case, when we have noise in both hops. After that, in Section \ref{section6}, we extend the results of section \ref{section5} for the model with random energy loss and in Section \ref{numer}, we provide some numerical results. Finally, we conclude our results in Section \ref{section7}.

\textbf{Notation}: We define $[1:M]=\{1,...,M\}$. Upper-case letters (e.g., $X$) denote Random Variables (RVs) and lower-case letters (e.g., $x$) their realizations. The probability mass function (p.m.f) of a RV $X$ with alphabet set $\Xc$ is denoted by $p_X(x)$; occasionally, subscript $X$ is omitted.
$X^j_i$ indicates a sequence of RVs $(X_i,X_{i+1},...,X_j)$; we use $X^j$ instead of $X^j_1$ for brevity.
The channel inputs at the transmitter and the relay are shown by $\{ {X_{1,1}},{X_{1,2}},{X_{1,3}},...\}$ and $\{ {X_{2,1}},{X_{2,2}},{X_{2,3}},...\}$, respectively. The output at the relay and the receiver's are denoted by $\{ {Y_{2,1}},{Y_{2,2}},{Y_{2,3}},...\}$ and $\{ {Y_{3,1}},{Y_{3,2}},{Y_{3,3}},...\}$, respectively.  $\pi_u$ denotes the steady state probability of the $u$-th state of a Markov chain. $A_\varepsilon ^{(n)}$ is the set of robust typical sequences \cite{Orlitsky} and $\tau _\varepsilon ^{(n)}$ is the set of weak typical sequences \cite{elGamal}.

\section{System Model}

\label{section2}

\begin{figure}
\centering
\includegraphics[width=9cm]{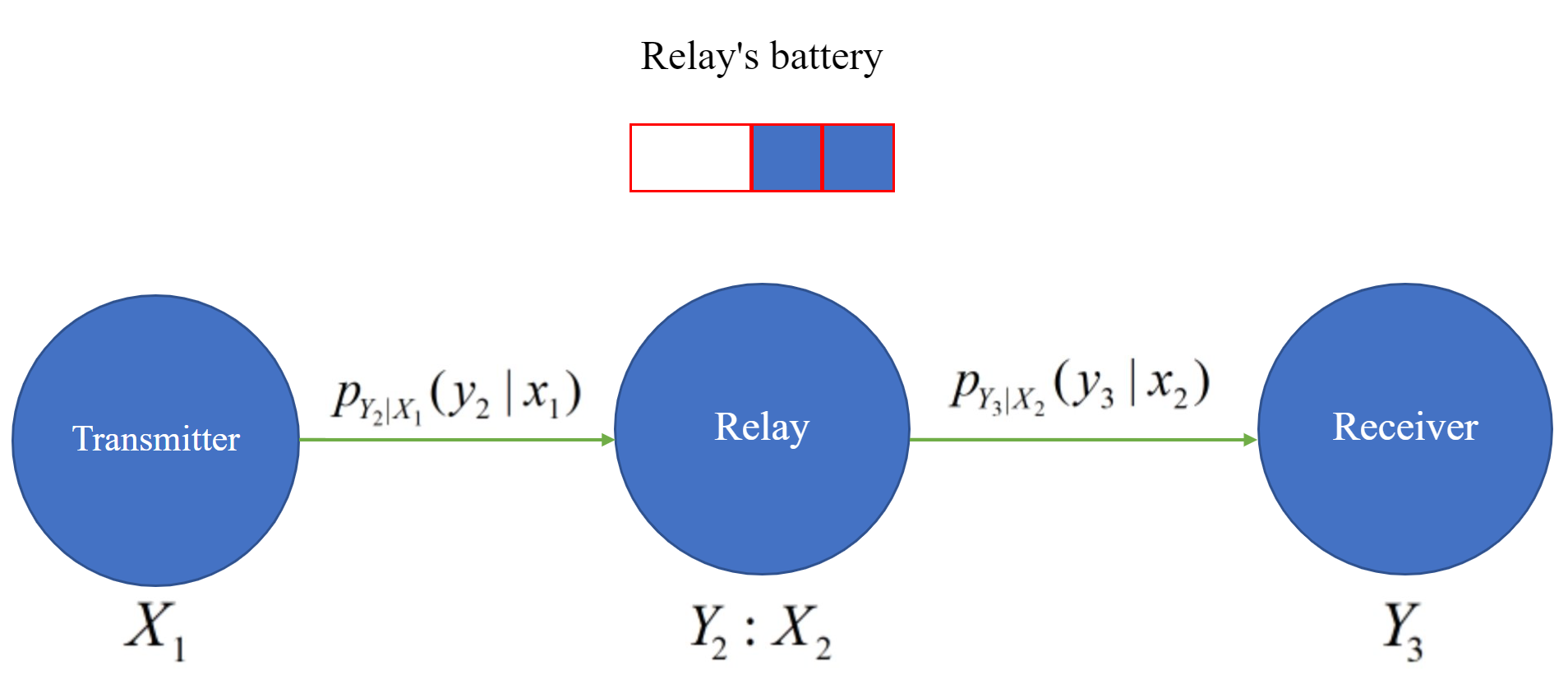}
\caption{Two-hop and energy harvesting relay channel with finite battery.}
\label{fig:model}
\end{figure}

We consider a binary two-hop relay channel illustrated in Fig. \ref{fig:model}, in which the relay node has energy restriction (i.e., the relay must harvest energy from its received signal to be able to transmit). Also, we assume a finite battery at the relay, which can save a finite number of energy units (up to $U$). Thus, the transmitted symbol depends on the harvested energy from the received symbols. We call the system a two-hop relay channel with a finite battery (THRC-FB).
The channel input at the transmitter and the channel outputs at the relay and the receiver have binary alphabets, i.e., $\mathcal{X}_1=\mathcal{Y}_2=\mathcal{Y}_3=\{0,1\}$. The alphabet of channel input at the relay is shown by $\mathcal{X}_2$ (will be introduced later). $\{ {U_1},{U_2},{U_3},...\} $ show the level of battery storage in the $i$-th transmission, which are modeled by the channel state and ${\cal U} =\{0,1,...,U\}$. $S_i$ are considered as: $S_1=(U_1,U_2),S_2=(U_2,U_3),S_3=(U_3,U_4),\ldots$.

In our system model transmitting symbol ``1" needs $m$ energy units, while the symbol ``0" can be sent with zero energy (the transmitting symbol "1" costs $m$ energy units for the relay). However, receiving the symbol "1" at the relay charges the battery with only one energy unit. This consideration shows the energy loss in the channel.
The transmitter does not have energy restriction and it can transmit any symbol in each channel use. The transmitter sends a message $M\in [1:{{2}^{nR}}]$ to the relay node in $n$ channel uses (by transmitting $X_1^n$). Then, the relay decodes this message and sends it to the receiver (by transmitting $X_2^n$).
The $\mathcal{X}_{2|u}$ is the set of symbols that can be transmitted by the relay, when the channel state is $u$, which shows the energy units stored in the relay's battery. $U$ is the maximum battery size.
Energy restriction at the relay is described as:
\begin{equation}
u < m \to {\mathcal{X} _{2|u}} \in \{ 0\},
\label{e1}
\end{equation}
\begin{equation}
u\ge m\to {{\mathcal{X} }_{2|u}}\in \{0,1\},
\label{e2}
\end{equation}
where $u \in [0:U]$. We assume that $m>1$, to consider the channel loss.

Encoding and decoding functions depend on battery size $U$, so we have to include this parameter in our code definition. A $({2^{nR}},n,U)$ code for  THRC-FB  consists of a message set $ [1:{{2}^{nR}}]$, an encoder function, which maps $m\in [1:{{2}^{nR}}]$ to $x_1^n(m)$, a set of relay encoder functions, which maps each past received sequence $y_2^{i - 1}$ to ${x_{{2,i}}}(y_2^{i - 1})$ for $i \in [1:n]$, and a decoder function, which estimates $\hat{m}$ from the received sequence $y_3^n$ at the receiver. We define the average probability of error as $P_e^{(n)} = P\{ M \ne \hat M\} $. A rate $R$ is achievable for  THRC-FB, if there exists a $({2^{nR}},n,U)$ code, for which we have $\mathop {\lim }\limits_{n \to \infty } P_e^{(n)} = 0$.

\section{Two-Hop Energy Harvesting Relay Channel With Finite Battery: Noisy Second Hop}
\label{section3}

In this section, we consider a binary memoryless channel between the relay and the receiver (the second hop). 
In this case, the the first hop is noise free, i.e., $Y_2=X_1$.
We consider that the memoryless channel between the relay and the receiver has the following conditional p.m.f:
\begin{equation}
\left\{ {\begin{array}{*{20}{c}}
{p({y_3} = 0\left| {{x_2} = 0} \right.) = {q_1}}\\
{p({y_3} = 1\left| {{x_2} = 1} \right.) =  {q_2}}
\end{array}}. \right.
\label{conditional p.m.f}
\end{equation}

\begin{figure}
\centering
\includegraphics[width=10cm]{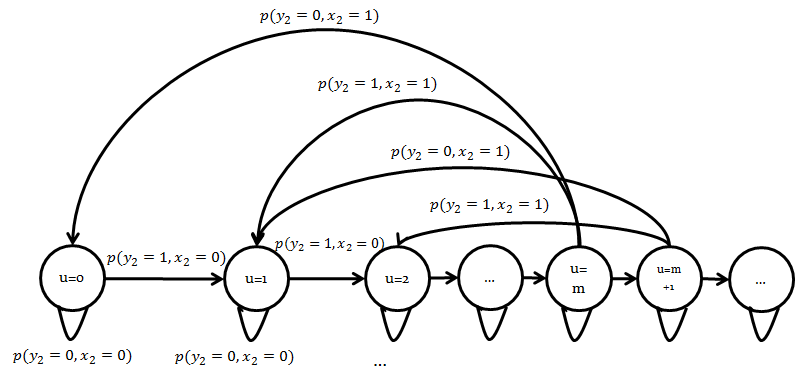}
\caption{State diagram of noiseless THRC-FB.}
 \label{fig:state}
\end{figure}

The main difficulty here is that the system has memory due to the energy restriction and finite battery size at the relay. Our approach is to model the energy units (stored in the relay's battery) as the state of the system. Also, we assume that the receiver does not know the state sequence, which is another challenge we face.
The state diagram of the channel is shown in Fig. \ref{fig:state}. As seen in Fig. \ref{fig:state}, when the battery is in state $u$ in the current transition, four cases can occur: i) if the relay receives symbol $1$ and transmits symbol $0$, then, the state in the next transition would be $u+1$ (except when $u=U$, where the next state will be the same); ii) if the relay receives symbol $0$ and transmits symbol $1$, then, the state in the next transition would be $u-m$ (in this case, $u \ge m$ is required, otherwise it does not occur); iii) if the relay receives symbol $1$ and transmits symbol $1$, then, the state in the next transition would be $u-m+1$ (in this case, $u\ge m$ is required, otherwise it does not occur); iv) if the relay receives symbol $0$ and transmits symbol $0$, then, the next state would be the same as the current state. The above discussions can be summarized as follows:
\begin{equation}
{U_{i + 1}} = \min \left\{ {{U_i} + {x_{1,i}} - m{x_{2,i}},U} \right\},
\label{model}
\end{equation}
where $x_{2,i}$ must satisfy (\ref{e1}) and (\ref{e2}).

In this case, the noise can change the transmitted symbol, and so its energy, randomly.
This means that if the channel converts symbol $``1"$ into symbol $``0"$, it would have zero energy and if the channel converts symbol $``0"$ into symbol $``1"$, its energy increases by  one energy unit. 
We call this system a two-hop relay channel with a finite battery (THRC-FB) with noisy second hop.

\subsection{Main result}

First, we state an achievability result for the THRC-FB with a noisy second hop.
Our achievability scheme is based on block Markov coding. For each state, we use superposition coding for codebook generation. The novelties  of our scheme is in the decoding method at the receiver without knowing the state sequence. We use backward and joint typicality decoding at the receiver.

\begin{theorem}
\label{theorem1}
The following rate, $R$, is achievable for THRC-FB with noisy second hop for the class of conditional p.m.fs, for which $q_1+q_2\ne1$:
\begin{equation}
\scalebox{1}[1]{$R < \mathop {\max }\limits_{p({x_{1\left| u \right.}},{x_{2\left| u \right.}})} \min \left\{ {\sum\limits_{u = 0}^U {{\pi _u}I({X_{2\left| u \right.}};{Y_{3\left| u \right.}})} ,\sum\limits_{u = 0}^U {{\pi _u}H({X_{1\left| u \right.}}\left| {{X_{2\left| u \right.}}} \right.)} } \right\}$},
\label{main theorem1}
\end{equation}
where $p(x_{1| u},x_{2| u})$ must be chosen such that for $u \ge m$, there is not any zero element in the p.m.f $p(x_{1| u},x_{2| u}),x_{1| u},x_{2| u}\in\{0,1\}$. In addition  for $u \in [0:m - 1]$, there is no zero element in the marginal p.m.f $p(x_{1| u}),x_{1| u}\in\{0,1\}$ and we must have $p({x_{2\left| u \right.}}) = \left\{ {\begin{array}{*{20}{c}}
{1\begin{array}{*{20}{c}}
{}&{{x_{2\left| u \right.}} = 0}
\end{array}}\\
{0\begin{array}{*{20}{c}}
{}&{{x_{2\left| u \right.}} = 1}
\end{array}}
\end{array}} \right.$.
\end{theorem}

\begin{IEEEproof}
Our scheme uses block Markov coding, where  $B$ blocks of transmissions (each of $n$ symbols) are sent to the relay node to transmit a sequence of $B-1$ independent and identically distributed (i.i.d.) messages ${m_b},b \in [1:B - 1]$, while the message of the last block ($m_B$) is deterministic. Similarly, the relay node sends $B$ blocks to the receiver, in which the message of the first block is deterministic and the messages of the remaining blocks are the same as the transmitter's message with one block delay. At the end of  each block, the relay decodes the message and sends it to the receiver in the next block. Since the state space is finite, we can control the initial state in each block with at most $U$ transmissions, i.e., we can enforce any desired initial state by changing the relay's battery. Thus, we assume that the initial state in each block can be adjusted and for simplicity, we do not contain these $U$ transmissions in our further discussions. In fact, by including these transmissions, each block contains $n+U$ channel uses instead of $n$. 

\textit{Codebook generation}:

For each state $u\in\mathcal{U}$, fix a p.m.f $p(x_{1| u},x_{2| u})$ satisfying the conditions described in Theorem \ref{theorem1}. For each state $u\in\mathcal{U}$ and for each block $b \in [1:B ]$, generate randomly and independently $2^{nR}$ sequences $x_{2| u }^{n_u+\delta }(m_{b-1})$, where $m_{b-1}\in [1:{{2}^{nR}}]$, each according to $\prod\limits_{i=1}^{n_u+\delta }{p_{X_{2| u }}({x_{2| u,i}})}$.
We map each $m_b \in [1,2^{nR}]$ to a vector $[{{m}_{0,b}},{{m}_{1,b}},...,{{m}_{U,b}}]$ where, $m_{u,b}\in [1,K_u]$, $K_u=2^{n{R_u}}$ and $R = \sum\limits_{u = 0}^U {{R_u}}$.  We note that $K_u$ is the size of each subcodebook for the state $u$.
For each ${m_{b-1}}\in [1:{{2}^{nR}}]$, we generate randomly and conditionally independently ${{K}_{u}}$ sequences $x_{1| u }^{n_u+\delta }(m_{u,b},m_{b-1})$, where $m_{u,b}\in [1:K_u]$, each according to $\prod\limits_{i=1}^{n_u+\delta }{{{p}_{{{X}_{1| u }}| {{X}_{2| u }} }}({{x}_{1| u ,i}}| {{x}_{2| u ,i}}(m_{b-1}) )}$, (note that $m_{u,b}$ is the submessage of the transmitter's subcodebook of state $u$ in block $b$). By convention, we have ${m_0} = {m_B} = 1$.
This codebook generation is shown in Fig.~\ref{fig:coding}.
In addition, we generate ${{2}^{nR}}$ i.i.d. random initial states $u(m_{b-1})$, where ${m_{b-1}}\in [1:{{2}^{nR}}]$, with p.m.f $\pi_u$. $u(m_{b-1})$ will be used as the initial state of channel in block $b$.

We remark that only the first $n_u$ bits (in each codeword) contain the message and we find $n_u$ for each state such that the  probability of error tends to zero. Other $\delta$ bits are generated to protect the channel's statistical properties from change. This means that we generate $\delta$ joint random bits from the p.m.f $p(x_{1| u},x_{2| u})$ for each message set {$(m_{u,b},m_{b-1})$}. Thus, if $n_u$ bits of codeword of state $u$ are sent completely before the codewords of other states, sending these $\delta$ bits would prevent the statistical properties of the channel state diagram from change. $\delta$ can be chosen as large as $n-\min(n_u)$ to satisfy the above condition. The transmission strategy is described in the following.

\begin{figure}
\centering
\includegraphics[width=7cm]{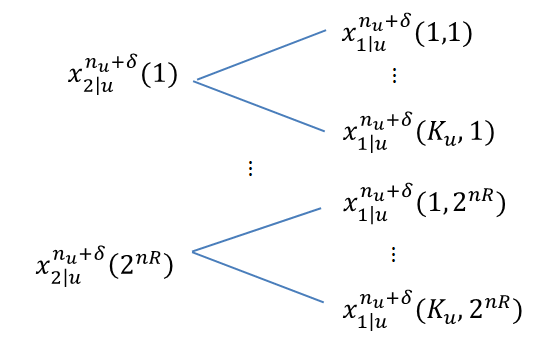}
\caption{The Superposition coding for each state with joint p.m.f $p({{x}_{1\left| u \right.}},{{x}_{2\left| u \right.}})$.}
 \label{fig:coding}
\end{figure}

\textit{Encoding (at the beginning of block $b$)}:

\textbf{Transmitter}:
To send message ${m_b}\in [1:\prod\limits_{u=0}^{U}{{{K}_{u}}}]$ in block $b$, the transmitter maps the message into a message vector $[{{m}_{0,b}},{{m}_{1,b}},...,{{m}_{U,b}}]$, ${{m}_{u,b}}\in [1:{{K}_{u}}]$. Then, knowing the message of previous block $m_{b-1}$, it selects the codeword $x_{1| u }^{n_u+\delta }(m_{u,b},m_{b-1})$ from the subcodebook corresponding to state $u$. In addition, it sets a vector as $[{l_{0,b}} = 1,{l_{1,b}} = 1,...,{l_{U,b}} = 1]$. For encoding in block $b$, the transmitter starts from the beginning of the block with $u_1$ as initial state. Then, in the $i$-th transmission, the transmitter sends ${x_{1\left| {u_i,{l_{u_i,b}}} \right.}}({m_{u_i,b}},{m_{b - 1}})$.  Encoding procedure is shown completely in Fig \ref{fig:flowchart}. Note that there is no noise between the transmitter and the relay, therefore, by the relay's encoding procedure (described in the next pragraph), the state sequence $U_i$ will bw known at the transmitter.

\begin{figure}
\centering
\includegraphics[width=6cm]{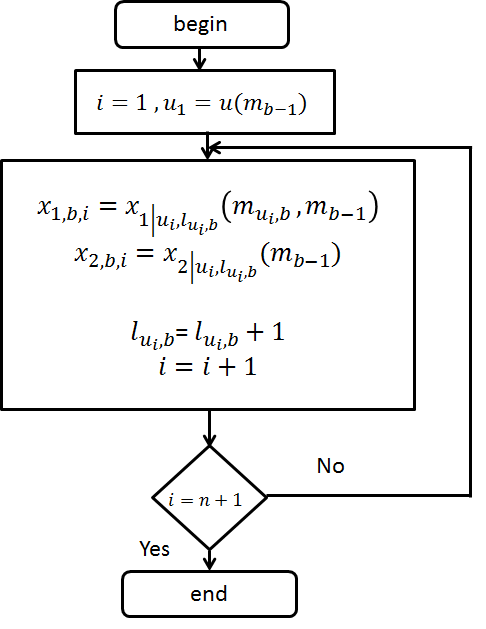}
\caption{The encoding procedure in the $b$-th block at the transmitter and the relay.}
 \label{fig:flowchart}
\end{figure}

\textbf{Relay:}
The relay sends message $m_{b-1}$ in  block $b$, so it selects codeword $m_{b-1}$ from each subcodebook of each state $u\in {\cal U}$ as the message of that subcodebook and chooses $u(m_{b-1})$ as the initial state (thus, the parameter $u_1$ in the previous paragraph is set to $u(m_{b-1})$). Then, in the $i$-th transmission, it sends ${x_{2\left| {u_i,{l_{u_i,b}}} \right.}}({m_{b - 1}})$ and sets ${l_{u_i,b}} = {l_{u_i,b}} + 1$. This procedure is shown in Fig. \ref{fig:flowchart}.

\textit{Decoding}:

\textbf{Relay:}
The state sequence in the block $b$, $u_b^n$, is known at the relay, because it knows its battery storage. Thus, for each state $u \in {\cal U}$, the relay makes a set ${A_u} = \{ i\left| {{U_i} = u} \right.\}$. If  $\left| {{A_u}} \right| \ge {n_u}$, the relay looks for an ${{\hat m}_{u,b}} = \{ {m_{u,b}}\left| {{x_{1\left| {u,k} \right.}}({m_{u,b}},m_{b-1}) = {x_{1,b,{i_k}}}} \right. , k \in [1:{n_u}] , {i_1} \le {i_2} \le ... \le {i_{{n_u}}}\} $. This procedure is shown in Fig. \ref{fig:flowchartdecode}.
In the analysis of the probability of error, we find the conditions that guarantee the ${{\hat m}_{u,b}}$ to be unique. Then, the relay forms the vector $[{{\hat m}_{0,b}},{{\hat m}_{1,b}},...,{{\hat m}_{U,b}}]$  and by the inverse of mapping used in encoding, it can decode $\hat{m}_b$.

\begin{figure}
\centering
\includegraphics[width=7.5cm]{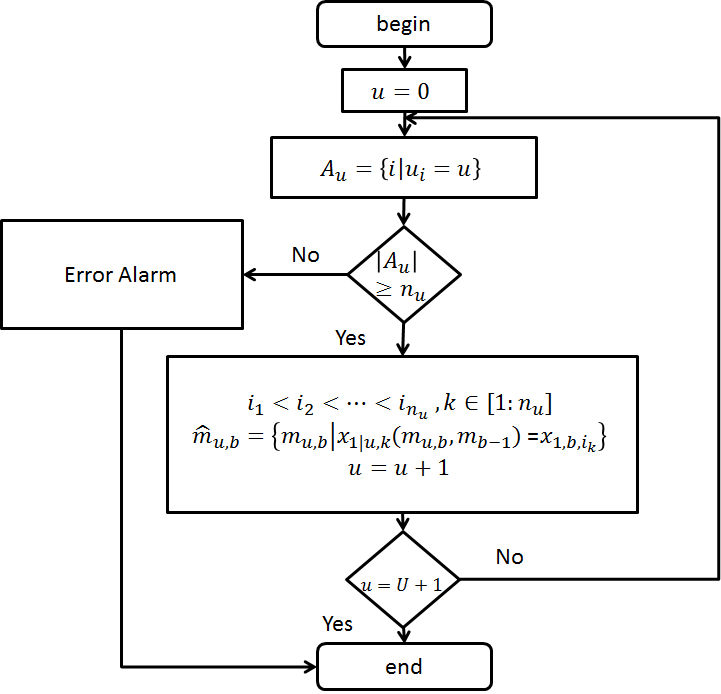}
\caption{Flowchart of decoding block $b$ in relay.}
 \label{fig:flowchartdecode}
\end{figure}

\textbf{Receiver}:
The receiver uses backward decoding. In the last block ($B$), the transmitted message from the transmitter is fixed, i.e., $m_B=1$. Assume that the relay transmits the message $m'_{B-1}$ in the last block, where $m'_{B-1}\in [1:{{2}^{nR}}]$. The receiver runs the flowchart shown in Fig. \ref{fig:flowchart} for each $m'_{B-1}$ and computes the sequence generated at the relay for each $m'_{B-1}$. We call these sequences $X_2^n(m_B = 1,m'_{B - 1})$. Then, the receiver looks for ${\hat{m}}_{B-1}$, which satisfies $(X_2^n(m_B = 1,{{\hat m}_{B - 1}}),Y_{3}^n(B)) \in \tau _\varepsilon ^{(n)}$, where  $Y_{3}^n(B)$ is the received sequence at the receiver in the block $B$.
To perform the above decoding, the receiver does not need the state sequence, because: 1) the codebooks and initial states are shared; 2) there is not any noise between the transmitter and the relay; 3) as seen in the flowchart of Fig. \ref{fig:flowchart}, since the transmitter's message (in  the last block $m_B=1$), the relay's message $m'_{B-1}$ and the initial state are known, after every transmission the next state can be determined and $X_2^n(m_B = 1,m'_{B - 1})$ could be derived for each $m'_{B - 1}\in [1:{{2}^{nR}}]$.
 When the message ${m}_{B-1}$ is decoded, the transmitter's message in block $B-1$ is known and the above procedure can be repeated to decode the previous blocks messages.

\textit{Analysis of the probability of error}:

Without loss of generality, we assume that the message $m_b=1$ is sent in block $b$, $b\in\{1,...,B\}$.
The probability of error is upper bounded by the sum of probabilities of error at the relay and the receiver.
The error events at the relay in each block are:
\begin{itemize}
\item ${\varepsilon ^{(1)}}=$ The relay does not receive the codeword of at least one of the codebooks (corresponding to a state $u$) completely. We define $\varepsilon _u^{(1)}$ as the event, in which the relay does not receive the codeword of the subcodebook related to the state $u$ completely.
\item ${\varepsilon ^{(2)}}=$ There are more than one  codeword equal to the received sequence in at least one of subcodebooks. We define $\varepsilon _u^{(2)}$ as the event, in which there are more than one codeword equal to the received sequence of the subcodebook related to the state $u$.
\end{itemize}
where it is seen that $p({\varepsilon ^{(i)}}) \le \sum\limits_u {p(\varepsilon _u^{(i)})} ,i \in \{ 1,2\}$.

First, we consider the ${\varepsilon ^{(1)}}$. We can see from (\ref{model}) and the way of choosing $p(x_{1| u},x_{2| u})$ that the Markov chain produced by the sequence $U_i$ is indecomposable. Now, we introduce a lemma, which states a sufficient condition for the existence of steady-state probabilities in an indecomposable finite-state Markov chain.

\begin{lemma}
\label{lem:steady}
Consider an indecomposable Markov chain with $r$ possible states. The steady-state probabilities exist, if there exists a state $\tilde{S}$ in the state diagram, which is accessible from itself in one transition (the probability of returning to itself in the next transition is nonzero).
\end{lemma}

\begin{IEEEproof}
The proof of this lemma is provided in Appendix \ref{appendix 1}.
\end{IEEEproof}

 We can conclude by Lemma~\ref{lem:steady} that the steady state probabilities exists for $U_i$, because we have $p(x_{1| u}=0,x_{2| u}=0) \ne 0$ by assumption for each $u\in[0:U]$. In addition, based on \cite[Theorem 6.6.3.]{ash}, in a finite Markov chain with steady state, the relative frequency of being in a state $u$ converges to the steady state probability ${\pi _u}$, in probability. Thus, if we choose ${n_u} = n({\pi _u} - \epsilon )$, the event ${\varepsilon ^{(1)}}$ does not occur with probability 1 and $p({\varepsilon ^{(1)}})$ goes to zero for large enough $n$.
Next, we provide a lemma to be used for bounding $p({\varepsilon ^{(2)}})$.

{
\begin{lemma}
\label{lemmaaepsilon}
Fix a joint p.m.f $p(u,x)$ and generate a random sequence $U^n$ according to $\prod\limits_{i = 1}^n {{p_U}({u_i})} $, then generate randomly and conditionally independently $2^{nR}$ sequences ${X^n}(m),m \in [1:{2^{nR}}]$, each according to $\prod\limits_{i = 1}^n {{p_{X\left| U \right.}}(\left. {{x_i}} \right|{u_i})}$. If we have $R < H(X\left| U \right.)$, then for a given $m$, probability of  the event $\bigcup\limits_{m' \ne m} {\{ {X^n}(m) = {X^n}(m')\} } $, would tend to zero.
\end{lemma}

\begin{IEEEproof}
The proof of this lemma is provided in Appendix \ref{appendix 2}.
\end{IEEEproof}
 }

Based on  Lemma \ref{lemmaaepsilon}, the probability of the second error event (${\varepsilon ^{(2)}}$) goes to zero, if for each $u\in[0:U]$, we have:
\begin{equation}\label{eqn:EE2}
\frac{{\log ({K_{u}})}}{{{n_u}}} < H({X_{1\left| u \right.}}\left| {{X_{2\left| u \right.}}} \right.) - \epsilon \ .
\end{equation}
If we substitute ${n_u} = n({\pi _u} - \epsilon )$ in \eqref{eqn:EE2}, we obtain:
\begin{equation}
{n_u} = n\left( {{\pi _u} - \epsilon } \right) \to {n_u} < n\left( {{\pi _u}} \right) \to \frac{1}{{{n_u}}} > \frac{1}{{n {{\pi _u}} }},
\label{eq29}
\end{equation}
and thus,
\begin{equation*}
R = \frac{{{{\log }_2}\mathop \prod \nolimits_{u = 0}^U {K_u}}}{n} = \mathop \sum \limits_{u = 0}^U \frac{{{\pi _u}{{\log }_2}{K_u}}}{{n{\pi _u}}}
\end{equation*}
\begin{equation}
< \mathop \sum \limits_{u = 0}^U \frac{{{\pi _u}{{\log }_2}{K_u}}}{{{n_u}}}.
\label{eq30}
\end{equation}
By  (\ref{eqn:EE2}) and (\ref{eq30}), we have:
\begin{equation}
\label{eq31}
R < \sum\limits_{u = 0}^U {{\pi _u}H({X_{1\left| u \right.}}\left| {{X_{2\left| u \right.}}} \right.)}.
\end{equation}

For analysis of the probability of error at the receiver, recall that in the last block, the transmitter's message is known, so, the receiver derives $X_2^n(m_B = 1,m^{'}_{B - 1})$, ${{{m}^{'}}_{B-1}}\in [1:{{2}^{nR}}]$. Since we assume that the relay sends $X_2^n(m_B = 1,m_{B - 1}=1)$,
if we have $(X_2^n(m_B = 1,{{m^{'}}_{B - 1}}),Y_{3}^n(B)) \in \tau _\varepsilon ^{(n)}$ for at least one $m^{'}_{B-1}\ne 1$, the error occurs at the receiver.


\begin{lemma}
For each $m'_{B - 1}\in [1:{{2}^{nR}}]$, the sequence  $X_2^n(m_B = 1,m'_{B - 1})$ is an independent regular Markov source.
\end{lemma}

\begin{IEEEproof}
The proof of this lemma is provided in Appendix \ref{appendix 3}.
\end{IEEEproof}

Since a regular Markov source is ergodic \cite[Theorem 6.6.2]{ash}, ${X_{2,i}}$ satisfies conditions of asymptotic equipartition property (A.E.P) Theorem \cite[Theorem 6.6.1]{ash}.
In this step of the proof, we prove the A.E.P theorem for the sequences $Y_3^n$ and joint sequences $(X_2^n,Y_3^n)$, when the sequence $X_2^n$ (which is a regular Markov source) has been affected by the memoryless channel (\ref{conditional p.m.f}) and the sequence $Y_3^n$ is generated. First of all, we assume that parameters $q_1$ and $q_2$ of the channel's conditional p.m.f are rational numbers and prove the A.E.P theorem for this class of channels. After that, we prove the A.E.P theorem for real $q_1$ and $q_2$.
First, we  introduce some definitions about finite-state channels \cite{ash}. Our purpose of mentioning the word ``state" in the previous sentence is completely different from $U_i$ (which we considered as the state up to here).
Our purpose of using  state here is to describe the model of finite-state channels introduced in \cite{ash} and use this model in studying a specific class of memoryless channels.

Consider the set of $S=\{s_1,...,s_r\}$ as the set of states, set of $\Gamma = \left\{ {{b_1},...,{b_t}} \right\}$ as the input alphabet and the set $\Delta $ as the output alphabet.
Consider $r \times r$ matrices $M_1,...,M_t$ as transition matrices (these matrices have nonnegative values with all row sums equal to one).
In addition, consider a function $g$, which assigns to each pair of $(b,s)$, $b\in \Gamma$, $s\in S$, a symbol $g(b,s) \in \Delta$.

Finite state channel works in this way; if the initial state is $s_j$ and an input $b_i$ is applied to the channel, a transition is made governed by the matrix $M_i$ (the probability of going to the state $s_k$ from the initial state $s_j$ is the element $M_i(s_j,s_k)$ in the $j$-th row and the $k$-th column of the matrix $M_i$). If the destination state is $s_k$, an output $g(b_i,s_k)$ is produced. Consider every finite product $M_{i(1)}M_{i(2)}...M_{i(k)},{i(1)},...,{i(k)} \in \{1,...,t\},k=\{1,2,...\}$, the finite state channel is called regular, if the steady state probabilities exist for the Markov chain determined by the transition matrix $M_{i(1)}M_{i(2)}...M_{i(k)}$.

\begin{lemma}
For rational $q_1$ and $q_2$, the memoryless channel given by conditional p.m.f (\ref{conditional p.m.f}) is a regular finite-state channel, so if a regular Markov source $X^n$ is its input and the output sequence $Y^n$ is generated, then we have:
\begin{equation}
 - \frac{1}{n}\log (p({Y^n})) \xrightarrow{prob.} H\{\underline{Y}\},
\label{a.e.p.rat}
\end{equation}
\begin{equation}
 - \frac{1}{n}\log (p({X^n,Y^n})) \xrightarrow{prob.} H\{\underline{X},\underline{Y}\}.
\label{a.e.p.rat2}
\end{equation}
\label{lemma4}
\end{lemma}
\begin{IEEEproof}
The proof of this lemma is provided in Appendix \ref{appendix 4}.
\end{IEEEproof}
Now, we prove  (\ref{a.e.p.rat}) and (\ref{a.e.p.rat2}) for real $q_1$ and $q_2$. For this purpose, we assume that the sequence $X^n$ (the input of the memoryless channel) is first passed from a memoryless channel in the form of (\ref{conditional p.m.f}) with parameters $q_1^k$ and $q_2^k$ with rational values, where $\mathop {\lim }\limits_{k \to \infty }q_1^k  = {q_1}$, $\mathop {\lim }\limits_{k \to \infty } q_2^k = {q_2}$ and $q_1^k+q_2^k\ne1, \forall k \in \{1,2,...\}$ (the reason of this inequality will be discussed later), and the sequence $\tilde Y_{[k]}^n$ is generated.  Then, $\tilde Y_{[k]}^n$ is passed from a memoryless channel in the form of (\ref{conditional p.m.f}) with parameters $\tilde q_1^k$ and $\tilde q_2^k$ with real values and the sequence $Y^n$ is generated.
 We have to examine that the channel between $\tilde Y_{[k]}$ and $Y$ exists or not. This statement is given in Lemma \ref{aditional channel}.

\begin{lemma}
The memoryless channel between $\tilde Y_{[k]}$ and $Y$ with parameters $0\le \tilde q_1^k \le 1$ and $0\le \tilde q_2^k \le 1$ exists for large enough $k$.
\label{aditional channel}
\end{lemma}

\begin{IEEEproof}
The proof of this lemma is provided in Appendix \ref{appendix 5}.
\end{IEEEproof}

We can show that:
\begin{equation}
\mathop {\lim }\limits_{k \to \infty } H({Y}\left| Y _{[k]}\right.) =\mathop {\lim }\limits_{k \to \infty } H({Y _{[k]}}\left| Y\right.) = 0.
\end{equation}

In this step, we introduce a number of Lemmas to prove A.E.P theorem for the  sequences $Y^n$ and $(X^n,Y^n)$ when $X^n$ is a regular Markov source. The structure of our proof is: for $Y^n$, we prove that $ - \frac{1}{n}\log (p({Y^n}))$ converges in probability to $ - \frac{1}{n}\log (p(\tilde Y_{[k]}^n,{Y^n}))$, $ - \frac{1}{n}\log (p(\tilde Y_{[k]}^n,{Y^n}))$ converges in probability to $ - \frac{1}{n}\log (p(\tilde Y_{[k]}^n))$, and $ - \frac{1}{n}\log (p(\tilde Y_{[k]}^n))$ conveges in probability to $H\{\underline{Y}\}$. For $(X^n,Y^n)$, we prove that $ - \frac{1}{n}\log (p({X^n},{Y^n}))$ converges in probability to $ - \frac{1}{n}\log (p({X^n},\tilde Y_{[k]}^n,{Y^n}))$,  $ - \frac{1}{n}\log (p({X^n},\tilde Y_{[k]}^n,{Y^n}))$ converges in probability to  $ - \frac{1}{n}\log (p({X^n},{\tilde Y_{[k]}^n}))$, and $ - \frac{1}{n}\log (p({X^n},{\tilde Y_{[k]}^n}))$ converges in probability to $H\{(\underline{X},\underline{Y})\}$. For this porpose, we have to describe some useful relations:

\begin{equation*}
{\left[ z \right]_ + } = \left\{ {\begin{array}{*{20}{c}}
{z,z \ge 0}\\
{0,z < 0}
\end{array}} \right.,
\end{equation*}
\begin{equation}
\left| z \right| = 2{\left[ z \right]_ + } - z,
\label{abs1}
\end{equation}
\begin{equation}
{\left[ {\log (z)} \right]_ + } \le z\frac{{\log e}}{e}.
\label{abs2}
\end{equation}

\begin{lemma}
$ - \frac{1}{n}\log (p({Y^n}))$ converges in probability to $ - \frac{1}{n}\log (p(\tilde Y_{[k]}^n,{Y^n}))$.
\label{lemma6}
\end{lemma}
\begin{IEEEproof}
The proof of this lemma is provided in Appendix \ref{appendix 6}.
\end{IEEEproof}

\begin{lemma}
$ - \frac{1}{n}\log (p(\tilde Y_{[k]}^n,{Y^n}))$ converges in probability to $ - \frac{1}{n}\log (p({\tilde Y_{[k]}^n}))$.
\label{lemma7}
\end{lemma}
\begin{IEEEproof}
The proof of this lemma is provided in Appendix \ref{appendix 7}.
\end{IEEEproof}

\begin{lemma}
$ - \frac{1}{n}\log (p({ Y^n}))$ converges in probability to $H\{\underline{Y}\}$.
\label{lemma8}
\end{lemma}
\begin{IEEEproof}
The proof of this lemma is provided in Appendix \ref{appendix 8}.
\end{IEEEproof}

Now, we prove the convergence of $ - \frac{1}{n}\log (p({X^n,Y^n}))$.
\begin{lemma}
$ - \frac{1}{n}\log (p({X^n,Y^n}))$ converges in probability to $ - \frac{1}{n}\log (p({X^n},\tilde Y_{[k]}^n,{Y^n}))$.
\label{lemma9}
\end{lemma}

\begin{IEEEproof}
The proof of this lemma is provided in Appendix \ref{appendix 9}.
\end{IEEEproof}

\begin{lemma}
$ - \frac{1}{n}\log (p({X^n,\tilde Y_{[k]}^n}))$ converges in probability to $ - \frac{1}{n}\log (p({X^n},\tilde Y_{[k]}^n,{Y^n}))$.
\label{lemma10}
\end{lemma}

\begin{IEEEproof}
The proof of this lemma is provided in Appendix \ref{appendix 10}.
\end{IEEEproof}

\begin{lemma}
$ - \frac{1}{n}\log (p({X^n, Y^n}))$ converges in probability to $H\{\underline{X},\underline{Y}\}$.
\label{lemma11}
\end{lemma}

\begin{IEEEproof}
The proof of this lemma is provided in Appendix \ref{appendix 11}.
\end{IEEEproof}

Now, we derive the probability of error at the receiver in block $B$ (last block). The error occurs if one of these events occurs:
\begin{itemize}
\item
$\scalebox{1}[1]{$\varepsilon _{B,r}^{(1)} =\left\{ (X_2^n({m_B} = 1,m{'_{B - 1}} = 1),Y_{3}^n(B)) \notin \tau _\varepsilon ^{(n)}\right\}$},$
\item
$\scalebox{1}[1]{$\varepsilon _{B,r}^{(2)} = \left\{\exists m{'_{B - 1}} \ne 1:(X_2^n({m_B} = 1,m{'_{B - 1}}),Y_{3}^n(B)) \in \tau _\varepsilon ^{(n)}\right\}$}.$

\end{itemize}

By Lemma \ref{lemma4},\ref{lemma8} and \ref{lemma11}, $p(\varepsilon _{B,r}^{(1)})$ goes to zero by tending $n$ to infinity. Then, we calculate $p(\varepsilon _{B,r}^{(2)})$ (for simplicity, the index of block number is omitted):
\begin{align}
p(\varepsilon _{B,r}^{(2)})&=\sum{\underset{( \underline{\alpha },\underline{\beta } )\in \tau_{\varepsilon }^{(n)}}{\mathop \sum }\,p( X_{2}^{n}=\underline{\alpha } )p( Y_{3}^{n}=\underline{\beta } )}\nonumber\\
&\scalebox{1}[1]{$\le {{2}^{-n\left( H\left\{ \underline{{{X}_{2}}} \right\}+\varepsilon  \right)}}{{2}^{-n\left( H\left\{ \underline{{{Y}_{3}}} \right\}+\varepsilon  \right)}}{{2}^{n\left( H\left\{ \underline{{{X}_{2}}},\underline{{{Y}_{3}}} \right\}-\varepsilon  \right)}}$}.
\end{align}

Thus, the upper bound on $p(\varepsilon _{B,r}^{(2)})$ is:
\begin{equation}
p(\varepsilon _{B,r}^{(2)})\le {{2}^{nR}}{{2}^{-n\left( H\left\{ \underline{{{X}_{2}}} \right\}+H\left\{ \underline{{{Y}_{3}}} \right\}-H\left\{ \underline{{{X}_{2}}},\underline{{{Y}_{3}}} \right\}-3\varepsilon  \right)}}.
\label{eq10}
\end{equation}

Thus, the probability of error tends to zero, if:
\begin{equation}
\scalebox{1}[1]{$R<H\left\{ \underline{{{X}_{2}}} \right\}+H\left\{ \underline{{{Y}_{3}}} \right\}-H\left\{ \underline{{{X}_{2}}},\underline{{{Y}_{3}}} \right\}=\underset{n\to \infty }{\mathop{\lim }}\,\frac{1}{n}\left( I\left( X_{2}^{n};Y_{3}^{n} \right) \right)$}.
\label{eq11}
\end{equation}

In addition, ${Y_3^n}$ is stationary, so we have:
\begin{equation*}
\scalebox{1}[1]{$\mathop {\lim }\limits_{n \to \infty } \frac{1}{n}H\left( {Y_3^n} \right) = \mathop {\lim }\limits_{n \to \infty } H({Y_{{3,n}}}\left| {{Y_{{3,{n - 1}}}},{Y_{{3,{n - 2}}}},...,{Y_{{3,1}}})} \right.$}.
\end{equation*}

The term on the left-hand side can be written as:
\begin{equation}
\scalebox{1}[1]{$H({Y_{{3,n}}}\left| {{Y_{{3,{n - 1}}}},...,{Y_{{3,1}}})} \right. \ge H({Y_{{3,n}}}\left| {{U_n},{Y_{{3,{n - 1}}}},...,{Y_{{3,1}}})} \right.$}
\label{eq12}
\end{equation}
\begin{equation}
= H({Y_{{3,n}}}\left| {{U_n}} \right.) = H({Y_{{3,1}}}\left| {{U_1}} \right.) = \sum\limits_{u = 0}^U {{\pi _u}H({Y_{3\left| u \right.}})}.
\label{eq13}
\end{equation}
The reasons for (\ref{eq13}) is that conditioned on $U_n$, the distribution of ${Y_{{3,n}}}$ is independent of ${Y_{{3,{n - 1}}}},...,{Y_{{3,1}}}$ and ${Y_{{3,n}}}$ is stationary.
On the other hand, we have a binary memoryless channel that satisfies:
\begin{equation*}
H\left( {Y_3^n\left| {X_2^n} \right.} \right) = \mathop \sum \limits_{k = 1}^n H\left( {{Y_{{3,k}}}\left| {{X_{{2,k}}}} \right.} \right)
\end{equation*}
for which we have:
\begin{align}
H\left( {{Y_{{3,k}}}\left| {{X_{{2,k}}}} \right.} \right)& = \mathop \sum \limits_x H\left( {{Y_{{3,k}}}\left| {{X_{{2,k}}} = x} \right.} \right)p\left( {{X_{{2,k}}} = x} \right)\nonumber\\
 &\scalebox{1}[1]{$= \mathop \sum \limits_x H\left( {{Y_{{3,k}}}\left| {{X_{{2,k}}} = x} \right.} \right)\mathop \sum \limits_u {\pi _u}p\left( {{X_{{2,k}}} = x\left| u \right.} \right)$}
\label{eq14}\\
 &\scalebox{1}[1]{$= \mathop \sum \limits_u {\pi _u}\mathop \sum \limits_x H\left( {{Y_3}\left| {{X_2} = x} \right.} \right)p\left( {{X_2} = x\left| u \right.} \right)$}
\label{e15}\\
& \scalebox{1}[1]{$= \mathop \sum \limits_u {\pi _u}H\left( {{Y_{3\left| u \right.}}\left| {{X_{2\left| u \right.}}} \right.} \right)$},
\label{eq16}
\end{align}
where (\ref{eq14}) is due to the law of total probability and (\ref{e15}) holds due to the stationarity of ${{X_{{2,k}}},{Y_{{3,k}}}}$. By combining (\ref{eq11}), (\ref{eq13}) and (\ref{eq16}), we derive:
\begin{equation}
\frac{1}{n}\left( {I\left( {X_2^n;Y_3^n} \right)} \right) \ge \mathop \sum \limits_u {\pi _u}I\left( {{Y_{3\left| u \right.}};{X_{2\left| u \right.}}} \right).
\label{eq17}
\end{equation}

Now, based on (\ref{eq10}) and (\ref{eq17}), we see that if  $R < \mathop \sum \limits_u {\pi _u}I\left( {{Y_{3\left| u \right.}};{X_{2\left| u \right.}}} \right)$,  the probability of error at the receiver will tend to zero. Hence the proof is complete.
\end{IEEEproof}

{Note that to find $K_u$, first we have to solve optimization problem in (\ref{main theorem1}), therefore, we can determine ${p({x_{1\left| u \right.}},{x_{2\left| u \right.}})}$. Next, we calculate $H\left( {{X_{1\left| u \right.}}\left| {{X_{2\left| u \right.}}} \right.} \right)$ and we consider ${K_u} < H\left( {{X_{1\left| u \right.}}\left| {{X_{2\left| u \right.}}} \right.} \right)$.

}
\section{Two-Hop Energy Harvesting Relay Channel With Finite Battery With Noise In The First Hop}
\label{section4}

In this section, we consider the THRC-FB with a noisy memoryless channel in the first hop, and we consider that the second hop is noiseless, so $Y_3=X_2$. Also, we assume a battery at the relay, which can save only $m$ energy units (required energy for transmission of only one energy-contained symbol, i.e., symbol ``1"), thus, $U=m$. We call this system THRC-FB with noisy first hop.

\begin{figure}
\centering
\includegraphics[width=5cm]{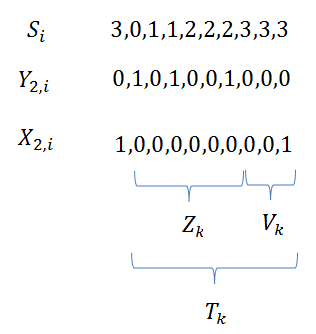}
\caption{An example for the relation between $Z_k$, $V_k$ and $T_k$, when $m=3$.}
 \label{timingfig}
\end{figure}

In this section, we adopt a completely different approach than what we used in section \ref{section3}. In this section, we model the channel between the relay and the receiver as a memoryless channel, so we have to introduce some additional notations. When the relay transmits its $(n-1)$-th symbol ``1", it has to wait to charge its battery by $m$ energy units. We show this waiting time by $Z_n$, therefore, ${Z_n} \in \{ m - 1 , m , m + 1,...\}$ is the number of transmissions between $(n-1)$-th relay's transmission of symbol ``1" and the first time, in which battery includes $m$ energy units after $(n-1)$-th relay's transmission of the symbol ``1" (note that $Z_n$ might become $m-1$ because the relay can receive a symbol with energy unit when it is transmitting $(n-1)$-th symbol ``1"). After that, the relay can decide to transmit its $n$-th symbol ``1" or wait for more, so ${V_n} \in \{ 1,2,...\}$ is the number of transmissions that the relay decides to wait for sending $n$-th symbol ``1" after the first time, in which the battery includes $m$ energy units proceeding the $(n-1)$-th relay's transmission of symbol ``1" and the relay can encode its message in the sequence $V_n$. Finally, $T_n$ is the number of transmissions between $(n-1)$-th and $n$-th relay's transmission of the symbol ``1". We would see:
\begin{equation}
\label{equivalent:ch}
T_n=V_n + Z_n.
\end{equation}
The above equation is shown in Fig. \ref{timingfig}.

\subsection{Related work}
The capacity of the noiseless energy harvesting channel with the unit-sized battery was determined by Tutuncuoglu et. al. \cite{ulukus}. In their system model, they considered an energy harvesting transmitter and an i.i.d. binary energy process available at the transmitter. Transmission of symbol ``1" needs one energy unit, while transmission of symbol ``0" does not need any energy. They considered an equivalent timing channel for the energy harvesting system to derive the capacity. The expression of the capacity is given by:

\begin{equation}
{C_T} = \mathop {\max }\limits_{p(u),v(u,z)} \frac{{I(U;T)}}{{E\{ T\} }},
\label{timming}
\end{equation}
where $U$ is an auxiliary variable and $v$ is a deterministic function.
The idea of the equivalent timing channel can be also applied to our two-hop relay channel model. We remark that we face a two-hop channel in comparison to the point-to-point channel of \cite{ulukus}. Our trick is to model the received sequence at the relay as an energy process to be able to adopt the timing channel approach.

\subsection{Main result}
In this section, we derive an achievable rate for the THRC-FB with noisy first hop. We use block Markov coding and backward decoding in our scheme. In addition, we use the equivalent timing channel for the second hop (the channel between the relay and the receiver). This means that the message is encoded in the number of waiting times between two transmitted symbols ``1" ($V_i$). This equivalent channel is a kind of memoryless channel with i.i.d. additive noise as described in (\ref{equivalent:ch}). In fact, after transmission of symbol ``1" in the $i$-th transmission at the relay, the relay's battery storage becomes empty until the first time relay receives $m$ energy units (this time is introduced as $Z_i$ in our system model and is a negative binomial RV with parameter $m$). As we see, the sequence $Z_i$ is i.i.d. additive noise in equivalent timing channel, which is casually known at the relay, hence, we can use coding scheme for the channel with casual state information at the transmitter.

\begin{theorem}
The following rate, $R$, is  achievable  for the THRC-FB with noisy first hop:
\begin{equation}
\label{mainthm}
\scalebox{1}[1]{$R < \mathop {\max }\limits_{p({x_1}),p(u),v(u,z)} \min \{ I({X_1};{Y_2}),\frac{{H(T)}}{{E\{ T\} }} - H({Y_2}\left| {{X_1}} \right.)\}.$}
\end{equation}
In this equation, $T \in \{m,m+1,...\}$ is the number of channel uses between two subsequent symbols ``1", $U$ is an auxiliary RV and $v$ is a deterministic function. $z$ is the timing noise imposed by the received sequence at the relay (this means that $z$ is the number of channel uses required for the relay to receive $m$ energy units and in this system is equal to the number of Bernoulli experiments with p.m.f $p(x_1)$ required for $m$ successes).

\end{theorem}

\begin{IEEEproof}
We use block Markov coding, in which $B$ blocks of transmissions (each of $n$ transmission) are sent to the relay. These $B$ blocks contain $B-1$ i.i.d. messages ${m_b},b \in [1:B - 1]$ and we have $m_0=m_B=1$. The relay sends these $B-1$ messages to the receiver with one block delay while the relay's message at the first block is fixed.
Now we describe our scheme in detail.

\textit{Codebook generation}:

 Fix p.m.f $p(u)$ and For each block $b \in [1:B]$, generate randomly and independently $2^{nR}$ sequences $u^l(m_{b-1})$ each according to $\prod\limits_{i = 1}^l {{p_U}({u_i})}$. Then, fix p.m.f $p(x_1)$ and for each $m_{b-1}$ and generate randomly and idependently $2^{nR}$ sequences $x_1^n({m_b,m_{b-1}})$ each according to $\prod\limits_{i = 1}^n {{p_{{X_1}}}({x_{1,i}})}$.  This codebook is used for encoding  at the transmitter. After that for each $x_1^n({m_b,m_{b-1}})$,  generate randomly and conditionally idependently ${2^{n\tilde R}}$ sequences ${{\tilde y}^n}({{\tilde m}_b},{m_b},{m_{b - 1}})$, each according to $\prod\limits_{i = 1}^n {{p_{{\tilde Y}\left| {{X_1}} \right.}}({{\tilde y}_i}\left| {{x_{1,i}({m_b},{m_{b - 1}})}} \right.)}$, where $p({\tilde y}\left| {{x_1}} \right.) = p({y_2}\left| {{x_1}} \right.)$. These sequences are used for decoding in receiver. We  also assume $n > \sum\limits_{i = 1}^l {{T_i}}$ for each $u^l$ (relation between $T_i$ and $u^l$ would be discussed in relay encoding). The codewords are shown in
Fig. \ref{fig:codebook_timing}.

\begin{figure}
\centering
\includegraphics[width=8cm]{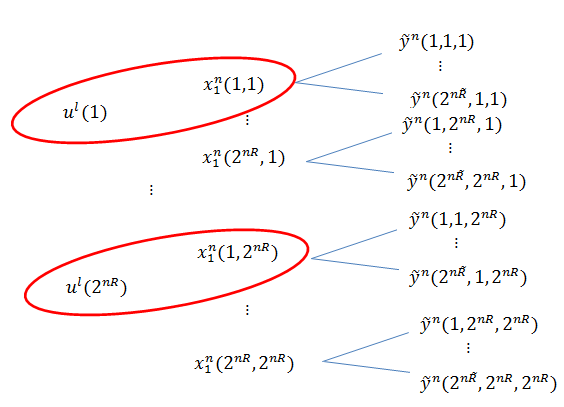}
\caption{Codebook generation for THRC-FB with noisy first hop.}
 \label{fig:codebook_timing}
\end{figure}

\textit{Encoding (at the beginning of block $b$)}:

\textbf{Transmitter:}
In the block $b$, transmitter's message is $m_b$ and also it knows $m_{b-1}$ (message of previous block), so transmitter sends $x_1^n({m_b},m_{b-1})$ ($m_0=1$ by convention).

\textbf{Relay:}
For encoding the relay's message, we model the channel between the relay and the receiver as a timing channel described before, thus, the second hop would become a channel with state, in which the channel state information is causally known at the relay, so we can use \cite[Theorem 7.2]{elGamal}. So, the timing noise sequence $Z^l$
is causally known at the relay and the relay would encode its message in the sequence $V^l$ and the receiver has to decode the relay's message with timing sequence $T^l$.

In block $b$, to send message $m_{b-1}$, the relay follows these steps:
\begin{enumerate}
\item
The relay picks $u^l(m_{b-1})$.
\item
For sending $i$-th symbol ``1" after $(i-1)$-th symbol ``1", the relay knows the sequences $T^l$ and $Y^n_2$ causally, so it can derive $Z_i$ ($Z_i$ is the timing noise for sending $i$-th symbol ``1", which is imposed by the received sequence at the relay).
\item
After $Z_i$, the relay waits for $v(u_i(m_{b-1}),Z_i)$ channel uses by sending symbol ``0". So, to send $i$-th symbol ``1" after $i-1$-th symbol ``1", the relay waits for $T_i=Z_i+v(u_i(m_{b-1}),Z_i)$ channel uses by sending symbol ``0".
\end{enumerate}

$Z_i$s are i.i.d. because we generate each bit of transmitter's mesages ($X_{1,i}({m_b,m_{b-1}})$) independently and identically distributed, {and due to memoryless channel ${Y_{1}}_{i_1}^{i_2}({m_b,m_{b-1}})$ and ${Y_{1}}_{i_2+1}^{i_3}({m_b,m_{b-1}})$ are independent} and the p.m.f of each $Z_i$ is equal to p.m.f of the RV defined as the number of i.i.d. Bernoulli experiments required for $m$ successes (negative binomial with parameter $m$). Note that $v$ is a deterministic function. Because we assume $n > \sum\limits_{i = 1}^l {{T_i}} $ for each $u^l$, after sending $l$-th symbol ``1", the relay sends symbol ``0" until the number of tranmissions becomes equal to $n$ in each block. So, we would consider $n = l(E\{ T\} + \delta ')$, where ${\delta '}$ could take every small value for large values of $n$ (note that by Markov inequality, we have $\mathop {\lim }\limits_{n \to \infty } p(l(E\{ T\} + \delta ') < \sum\limits_{i = 1}^l {{T_i}} ) = 0$).

\textit{Decoding}:

\textbf{Relay:}
To decode the message $m_b$ at the end of block $b$ the relay, knowing $m_{b-1}$, looks for a unique $\hat{m_b}$, which satisfies $(x_1^n({{\hat m}_b},m_{b-1}),Y_2^n(b)) \in \tau_\varepsilon ^{(n)}\left(p(x_1,y_2)\right)$.

\textbf{Receiver:}
We use backward decoding at the receiver. The receiver knows that in the last block, the transmitter's message is fixed, i.e., $m_B=1$.
For decoding message $m_{B-1}$ in block $B$, the receiver considers each ${m_{B - 1}} \in [1:{2^{nR}}]$ as the message of block $B-1$ and for each ${m_{B - 1}}$, it considers the sequences ${{\tilde y}^n}({{\tilde m}_B},{m_B} = 1,{m_{B - 1}}),{{\tilde m}_B} \in [1:{2^{n\tilde R}}]$ as the received sequence at the relay ($Y^n_2(B)$) and for each of them, it determines ${Z^l}({{\tilde m}_B},{m_B} = 1,{m_{B - 1}})$ and ${T^l}({{\tilde m}_B},{m_B} = 1,{m_{B - 1}})$ (this notation means that ${Z^l}({{\tilde m}_B},{m_B} = 1,{m_{B - 1}})$ would be the timing noise and ${T^l}({{\tilde m}_B},{m_B} = 1,{m_{B - 1}})$ would be the received timing sequence at the receiver, if ${{\tilde y}^n}({{\tilde m}_B},{m_B} = 1,{m_{B - 1}})$ is the received sequence at the relay and the relay's message is $m_{B-1}$). This process is feasible because: ${Z^l}({{\tilde m}_B},{m_B} = 1,{m_{B - 1}})$ could be determined by ${{\tilde y}^n}({{\tilde m}_B},{m_B} = 1,{m_{B - 1}})$ (considered as received sequence at the relay in block $B$) and $u^l(m_{B-1})$, because by these two sequences, timing noise could be determined and ${T^l}({{\tilde m}_B},{m_B} = 1,{m_{B - 1}})$ could be determined by $m_{B - 1}$ and the equation $T_i=Z_i+v(u_i(m_{B-1}),Z_i)$. After that, the receiver looks for a unique $(\hat m_{B-1})$, which satisfies $\exists {\tilde m_B} \in [1:{2^{n\tilde{R}}}]:\left({T^l}({\tilde m_B},{m_B} = 1,{\hat m_{B - 1}}) , {T^l}(B)\right)\in \tau_\varepsilon^{(n)} \left(p_T(t,t')\right)$ ($T^l(B)$ is the real timing sequence received in the block $B$ in the receiver and $p_T(t,t')$ denotes a joint distribution, in which $p_T(t,t)=1,t\in \{m,m+1,...\}$,$p_T(t,t')=0,t\ne t'$ and the marginal distribution $\sum\limits_{t'} {{p_T}(t,t')}$ is equal to the distribution of timing sequence $T$). By determining $\hat{m}_{B-1}$, we repeat this procedure for the previous blocks to discover $m_b,b \in [1,B-1]$.

\textit{Analysis of the probability of error}:

Without loss of generality, we assume that $m_b=1$ is sent in block $b \in [1:B]$. We define the following error events for each block $b$:
\begin{equation*}
\scalebox{1}[1]{${\varepsilon _1}(b) = \{ (X_1^n({{\hat m}_b} = 1, m_{b-1}=1),Y_2^n(b)) \notin \tau_\varepsilon ^{(n)}(p(x_1,y_2))\}$},
\end{equation*}
\begin{equation*}
\scalebox{1}[1]{${\varepsilon _2}(b) = \{ (X_1^n({{\hat m}_b} \ne 1,m_{b-1}=1),Y_2^n(b)) \in \tau_\varepsilon ^{(n)}(p(x_1,y_2))\}$},
\end{equation*}
\begin{equation*}
\scalebox{1}[1]{${\varepsilon _3}(b) = \{ \forall {{\tilde m}_{b+1}} \in [1:{2^{n\tilde R}}]:\left({T^l}({{\tilde m}_{b+1}},{m_{b+1}} = 1,{{\hat m}_{b }} = 1) , {T^l}({b+1})\right) \notin \tau_\varepsilon^{(n)} \left(p_T(t,t')\right)\}$},
\end{equation*}
\begin{equation*}
\scalebox{1}[1]{${\varepsilon _4}(b) = \{\exists {{\hat m}_{b}} \ne 1, \exists {{\tilde m}_{b+1}} \in [1:{2^{n\tilde R}}]:\left({T^l}({{\tilde m}_{b+1}},{m_{b+1}} = 1,{{\hat m}_{b}} ) , {T^l}(b+1)\right) \in \tau_\varepsilon^{(n)} \left(p_T(t,t')\right)\}$}.
\end{equation*}
We can see:
\begin{equation*}
\scalebox{1}[1]{$p(\varepsilon (b)) = p({{\hat m}_b} \ne 1) \le p({\varepsilon _1}(b)) + p({\varepsilon _2}(b)) + p({\varepsilon _3}(b)) + p({\varepsilon _4}(b))$}.
\end{equation*}
As $n \to \infty$, $p({\varepsilon _1}(b))$ tends to zero by law of large numbers (L.L.N). Based on joint A.E.P Theorem \cite[Theorem 7.6.1]{cover}, if $R <   I({X_1};{Y_2})$, $p({\varepsilon _2}(b))$ tends to zero.  Next, we show in Lemma  \ref{lemmaaepsilon2} that, if  $\tilde R > H({Y_2}\left| {{X_1}} \right.)$, with probability equal to one, there would exist a sequence ${{\tilde y}^n}({{\tilde m}_{b+1}},{m_{b+1}} = 1,{m_{b}=1})$ such that ${{\tilde y}^n}({{\tilde m}_{b+1}},{m_{b+1}} = 1,{m_{b}} = 1) = Y_2^n(b+1)$. So, there exists at least one ${T^l}({{\tilde m}_{b+1}},{m_{b+1}} = 1,{{\hat m}_{b }} = 1)$ such that 
$({T^l}({{\tilde m}_{b+1}},{m_{b+1}} = 1,{{\hat m}_{b }} = 1),T^l(b+1))\in \tau_\varepsilon^{(n)} \left(p_T(t,t')\right)$. Therefore $p(\varepsilon _3(b))$ tends to zero.
\begin{lemma}
\label{lemmaaepsilon2}
Fix a joint p.m.f $p(u,x)$ and generate a random sequence $U^n$ according to $\prod\limits_{i = 1}^n {{p_U}({u_i})} $, then generate randomly and conditionally independently $2^{nR}$ sequences ${X^n}(m),m \in [1:{2^{nR}}]$, each according to $\prod\limits_{i = 1}^n {{p_{X\left| U \right.}}(\left. {{x_i}} \right|{u_i})}$. If we have $R > H(X\left| U \right.)$, then for a given $m$, probability of  the event $\bigcup\limits_{m' \ne m} {\{ {X^n}(m) = {X^n}(m')\} } $, would tend to one.
\end{lemma}
\begin{IEEEproof}
The proof of this lemma is provided in Appendix \ref{appendix 12}.
\end{IEEEproof}
By our coding scheme, we can see that for the message set $(m_{b+1}=1,m_b=1)$, we use $u^l(m_b=1)$ and $x^n_1(m_{b+1}=1,m_b=1)$ and for the message set  $(m_{b+1}=1,m_b \ne 1)$, we use $u^l(m_b \ne 1)$ and $x^n_1(m_{b+1}=1,m_b \ne 1)$, this procedure is shown by red lines in Fig.  \ref{fig:codebook_timing}. So, ${T^l}({{\tilde m}_{b+1}},{m_{b+1}} = 1,{{\hat m}_{b }} = 1)$ and  ${T^l}({{\tilde{\tilde m}}_{b+1}},{m_{b+1}} = 1,{{\hat m}_{b }} \ne 1)$ would be independent and identical distributed sequences for each pair of $\tilde m _{b+1}$ and $\tilde{\tilde m}_{b+1}$. So,  the probability of the event $\left({T^l}({{\tilde m}_{b+1}},{m_{b+1}} = 1,{{\hat m}_{b}} \ne 1) , {T^l}(b+1)\right) \in \tau_\varepsilon^{(n)} \left(p_T(t,t')\right)$ would be upper bounded by ${2^{ - l(H(T) - \varepsilon) }}$ based on joint A.E.P Theorem \cite[Theorem 7.6.1]{cover}. So, we would have:
\begin{equation*}
\scalebox{1}[1]{$p({\varepsilon _4}(b)) \le {2^{n(R + \tilde R)}}{2^{ - l(H(T) - \varepsilon )}} = {2^{l(E\{ T\}  + \delta ')(R + \tilde R)}}{2^{ - l(H(T) - \varepsilon) }} $}
\end{equation*}
\begin{equation*}
 = {2^{l(E\{ T\}  + \delta ')(R + \tilde R - \frac{{H(T)}}{{E\{ T\}  + \delta '}}+ \varepsilon)  }}.
\end{equation*}
Thus, by $l \to \infty$, $p(\varepsilon_4(b))$ tends to zero, if $R + \tilde R < \frac{{H(T)}}{{E\{ T\} }}$, hence we can derive the bound $R < \frac{{H(T)}}{{E\{ T\} }} - H({Y_2}\left| {{X_1}} \right.)$. Also $p({\varepsilon _3}(b))$ tends to zero by L.L.N, as $n \to \infty$.
\end{IEEEproof}

\section{Two-Hop Energy Harvesting Relay Channel With Finite Battery With Noise In Both Hops}
\label{section5}
In this section, we extend the results of section \ref{section3} to the model of a two-hop energy harvesting relay channel with finite battery with noise in both hops. We consider memoryless channels in both hops. We call this model THRC-FB with noisy both hops. In this section, we assume $U\ge m$ and our notations are the same as section \ref{section3}.

\subsection{Main result}

Now, we provide an achievability theorem for THRC-FB with noisy both hops. Our coding approach is the same as section \ref{section3}.

\begin{theorem}
\label{theorem3}
The following rate, $R$, is achievable for THRC-FB with noisy both hops for the class of conditional p.m.fs, for which $q_1+q_2\ne1$ in the second hop (see (\ref{conditional p.m.f})):
\begin{equation}
\scalebox{.78}[1]{$R < \mathop {\max }\limits_{p({x_1}),p({x_{2\left| u \right.}})} {\mkern 1mu} \min \{ \sum\limits_{u = 0}^U {{\pi _u}I({X_{2\left| u \right.}};{Y_{3\left| u \right.}})}  - H({Y_2}\left| {{X_1}} \right.),I({X_1};{Y_2})\},$}
\label{main theorem3}
\end{equation}
where $p(x_{1})$ and $p(x_{2| u})$ must be chosen such that for $u \ge m$, there is not any zero element in the p.m.f $p(x_{1})p(x_{2| u})$. In addition,  for $u \in [0:m - 1]$, we must have $p({x_{2\left| u \right.}}) = \left\{ {\begin{array}{*{20}{c}}
{1\begin{array}{*{20}{c}}
{}&{{x_{2\left| u \right.}} = 0}
\end{array}}\\
{0\begin{array}{*{20}{c}}
{}&{{x_{2\left| u \right.}} = 1}
\end{array}}
\end{array}} \right.$.
\end{theorem}

\textit{Remark}: Maximization in  (\ref{main theorem3}) is over p.m.fs $p(x_1)$ and $p(x_{2|u})$, while maximization in  (\ref{main theorem1}) is over joint p.m.fs $p(x_{1|u},x_{2|u})$, therefore, the domain of maximization in  (\ref{main theorem3}) is much smaller than that of (\ref{main theorem1}).

\begin{IEEEproof}
The same as previous theorems, we use block Markov coding.

\textit{Codebook generation}:

For each state $u$,
and for each block $b \in [1:B]$, generate randomly and independently $2^{nR}$ sequences  $x_{2| u }^{n_u+\delta }(m_{b-1})$ each according to$\prod\limits_{i=1}^{n_u+\delta }{p_{X_{2| u }}({x_{2| u,i}})}$, for which we have $m_{b-1}\in [1:{{2}^{nR}}]$.
Then, for each  $m_{b-1}$, generate randomly and independently $2^{nR}$ sequences $x_1^n({m_b,m_{b-1}})$, each according to $\prod\limits_{i = 1}^n {{p_{{X_1}}}({x_{1,i}})}$. After that, for each sequence  $x_1^n({m_b,m_{b-1}})$, generate randomly and conditionally independently ${2^{n\tilde R}}$ sequences  ${{\tilde y}^n}({{\tilde m}_b},{m_b},{m_{b - 1}})$, each according to  $\prod\limits_{i = 1}^n {{p_{{\tilde Y}\left| {{X_1}} \right.}}({{\tilde y}_i}\left| {{x_{1,i}({m_b},{m_{b - 1}})}} \right.)}$, for which we have $p({\tilde y}\left| {{x_1}} \right.) = p({y_2}\left| {{x_1}} \right.)$. In addition, generate ${{2}^{nR}}$ random initial states $u(m_{b-1})$ for each  $b\in[1:B]$, each according to the p.m.f $\pi_u$, for which  ${m_{b-1}}\in [1:{{2}^{nR}}]$, and we use them as the initial state in block $b$. note that the same as Theorem  \ref{theorem1}, only the first  $n_u$ bits of each code in the codebook related to the state $u$ contains message, and other $\delta$ bits are generated to prevent channels statistical properties from change, so we can choose $\delta$ as large as $n-min(n_u)$. This kind of codebook generation is illustrated in Fig \ref{fig:random_energy_codebook}.

\begin{figure}
\centering
\includegraphics[width=8cm]{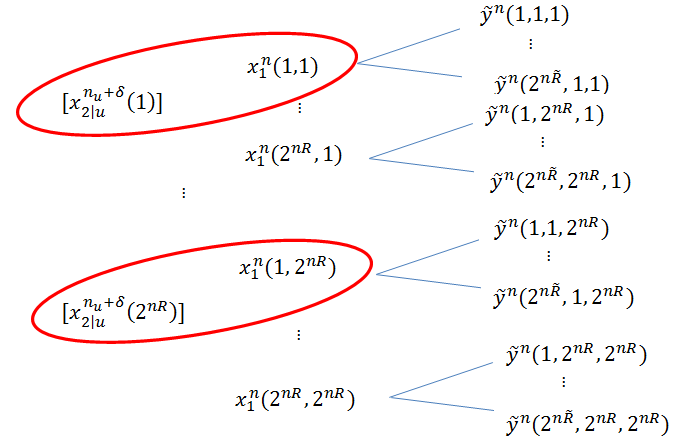}
\caption{Codebook generation for THRC-FB with noisy both hops.}
 \label{fig:random_energy_codebook}
\end{figure}

\textit{Encoding (at the beginning of block $b$)}:

\textbf{Transmitter:}
In block $b$, the transmitter's message is $m_b$ and the message of previous block is $m_{b-1}$, so, the transmitter sends the sequence $x_1^n({m_b},m_{b-1})$ (by the convention, we have $m_0=1$)

\textbf{Relay:}
The relays encoding strategy is the same as the strategy introduced in Theorem
\ref{theorem1}.
Note that for adjusting the initial state in each block, the transmitter can send $n\delta'$ number of symbol $"1"$ at the beginning of each block, by taking large values for $n$, if we have $p({y_2} = 1\left| {{x_1} = 1} \right.) \ne 0$, then, with probability equal to one, more than $U$ number of symbol $"1"$ would be received at the relay (because $U$ is a bounded number), so, the relay can take any arbitrary value for the initial state. By taking small enough $\delta'$, the transmission rate would not change.

\textit{Decoding}:

\textbf{Relay:}
To decode $m_b$ at the end of block $b$, the relay knows $m_{b-1}$, so, it looks for a unique $\hat{m_b}$, for which we have
$(x_1^n({{\hat m}_b},m_{b-1}),Y_2^n(b)) \in \tau_\varepsilon ^{(n)}$, where $Y_2^n(b)$ is the received sequence at the relay in block $b$.

\textbf{Receiver:}
The receiver uses backward decoding.
The receiver knows that $m_B=1$, so, for decoding $m_{B-1}$, the receiver considers each ${m_{B - 1}} \in [1:{2^{nR}}]$ as the message of  block $(B-1)$ and for each of them, it considers ${2^{n \tilde R}}$ sequences ${{\tilde y}^n}({{\tilde m}_B},{m_B} = 1,{m_{B - 1}}),{{\tilde m}_B} \in [1:{2^{n\tilde R}}]$ as the received sequence at the relay, then, it runs the flowchart in Fig \ref{fig:flowchart} and derives the sequence, which relay would generate for each of them. We call these sequences ${X^n_2}({{\tilde m}_B},{m_B} = 1,{m_{B - 1}})$. After that, the receiver looks for a unique
$\hat m_{B-1}$, for which we have
$\exists {{\tilde m}_B} \in [1:{2^{n\tilde{R}}}]:(X_2^n({{\tilde m}_B},{m_B} = 1,{{\hat m}_{B - 1}}),Y_3^n(B)) \in \tau_\varepsilon ^{(n)}$, where
$Y_3^n(B)$ is the received sequence at the receiver in block $B$. By determining $\hat{m}_{B-1}$, the receiver repeats this procedure for the previous blocks , thus,
$m_b,b \in [1,B-1]$ would be determined.

\textit{Analysis of the probability of error}:

Without loss of generality, we assume that for each $b \in [1:B]$, we have $m_b=1$.
We define the following error events for each block:
\begin{equation*}
{\varepsilon _1}(b) = \{ (X_1^n({{\hat m}_b} = 1, m_{b-1}=1),Y_2^n(b)) \notin \tau_\varepsilon ^{(n)}\},
\end{equation*}
\begin{equation*}
{\varepsilon _2}(b) = \{ (X_1^n({{\hat m}_b} \ne 1,m_{b-1}=1),Y_2^n(b)) \in \tau_\varepsilon ^{(n)}\},
\end{equation*}
\begin{equation*}
\scalebox{1}[1]{${\varepsilon _3}(b) = \{ \forall {\tilde m_{b + 1}} \in [1:{2^{n\tilde R}}]:(X_2^n({\tilde m_{b + 1}},{m_{b + 1}} = 1,{\hat m_b} = 1),Y_3^n(b + 1)) \notin \tau_\varepsilon ^{(n)}\} $},
\end{equation*}
\begin{equation*}
\scalebox{1}[1]{${\varepsilon _4}(b) = \{ \exists {\tilde m_{b + 1}} \in [1:{2^{n\tilde R}}],\exists \hat m_b \ne 1:(X_2^n({\tilde m_{b + 1}},{m_{b + 1}} = 1,{\hat m_b} ),Y_3^n(b + 1)) \in \tau_\varepsilon ^{(n)}\}$},
\end{equation*}
 so, we can see:
\begin{equation*}
\scalebox{1}[1]{$p(\varepsilon (b)) = p({{\hat m}_b} \ne 1) \le p({\varepsilon _1}(b)) + p({\varepsilon _2}(b)) + p({\varepsilon _3}(b)) + p({\varepsilon _4}(b))$}.
\end{equation*}

Based on L.L.N, 
$p({\varepsilon _1}(b))$ tends to zero as $n$ tends to infinity. In addition, based on the joint A.E.P Theorem \cite[Theorem 7.6.1]{cover}, if $R < I({Y_2}; {{X_1}})$, then $p({\varepsilon _2}(b))$ tends to zero. Based on Lemma \ref{lemmaaepsilon2}, if
$\tilde R > H({Y_2}\left| {{X_1}} \right.)$, then, there exists a sequence ${{\tilde y}^n}({{\tilde m}_{b+1}},{m_{b+1}} = 1,{m_{b}=1})$ such that 
${{\tilde y}^n}({{\tilde m}_{b+1}},{m_{b+1}} = 1,{m_{b}} = 1) = Y_2^n(b+1)$, so, there exists a sequence   $X_2^n({{\tilde m}_{b + 1}},{m_{b + 1}} = 1,{{\hat m}_b} = 1)$, for which we have
$(X_2^n({{\tilde m}_{b + 1}},{m_{b + 1}} = 1,{{\hat m}_b} = 1),Y_3^n(b + 1)) \in \tau_\varepsilon ^{(n)}$ (because of Lemmas \ref{lemma8} and \ref{lemma11}). Thus, $p(\varepsilon _3(b))$ tends to zero.

At last, for $\forall {{\tilde m}_{b + 1}}\in [1:2^{n\tilde{R}}]$, the pair of sequences  $X_2^n({{\tilde m}_{b + 1}},{m_{b + 1}} = 1,{{\hat m}_b} \ne 1) $ and $Y_3^n(b+1)$ are independent, so,  similar to the argument given in the proof of Theorem \ref{theorem1}, if
$R + \tilde R < \sum\limits_{u = 0}^U {{\pi _u}I({X_{2\left| u \right.}};{Y_{3\left| u \right.}})}  \le \mathop {\lim }\limits_{n \to \infty } {\mkern 1mu} \frac{1}{n}\left( {I\left( {X_2^n;Y_3^n} \right)} \right)$, then $p(\varepsilon _4(b))$ tends to zero, thus, we must have $R < \sum\limits_{u = 0}^U {{\pi _u}I({X_{2\left| u \right.}};{Y_{3\left| u \right.}})}  - H({Y_2}\left| {{X_1}} \right.)$.
\end{IEEEproof}

\section{Two-Hop Energy Harvesting Relay Channel With Finite Battery With Noise In Both Hops And Random Energy Loss}
\label{section6}
Until the previous section, we had deterministic energy loss and this means that if a symbol $"1"$ is received at a node, only one energy unit could be extracted from it.
In this section, we extend the result of the previous section to the same model with random energy loss. Random energy loss means that if a symbol $"1"$ is received at a node, then the extracted energy units will be a random number between $0$ and $m-1$, hence, if we show the number of energy units by $E$, then $E\in [0:m-1]$. We can see that the form of the channel state diagram will change in this case.  We call this model THRC-FB with noisy both hops and random energy loss. Note that in this model, $X_1$, $Y_2$ and $E$ makes a Markov chain.

\subsection{Main result}
Now, we extend Theorem \ref{theorem3} to THRC-FB with noisy both hops and random energy loss.

\begin{theorem}
\label{theorem4}
The following rate, $R$, is achievable for THRC-FB with noisy both hops and random energy loss for the class of conditional p.m.fs, for which $q_1+q_2\ne1$ in the second hop (see (\ref{conditional p.m.f})):
\begin{equation}
\scalebox{1}[1]{$R < \mathop {\max }\limits_{p({x_1}),p({x_{2\left| u \right.}})} {\mkern 1mu} \min \{ \sum\limits_{u = 0}^U {{\pi _u}I({X_{2\left| u \right.}};{Y_{3\left| u \right.}})}  - H({E}\left| {{X_1}} \right.),I({X_1};{Y_2})\},$}
\label{main theorem4}
\end{equation}
where $p(x_{1})$ and $p(x_{2| u})$ must be chosen such that the conditions of Lemma \ref{lem:steady} is satisfied (the state diagram is indecomposable and there exists at least one state, for which we can stay in it in one transition). In addition,  for $u \in [0:m - 1]$, we must have $p({x_{2\left| u \right.}}) = \left\{ {\begin{array}{*{20}{c}}
{1\begin{array}{*{20}{c}}
{}&{{x_{2\left| u \right.}} = 0}
\end{array}}\\
{0\begin{array}{*{20}{c}}
{}&{{x_{2\left| u \right.}} = 1}
\end{array}}
\end{array}} \right.$.
\end{theorem}

\begin{IEEEproof}
Proof of this theorem is the same as Theorem \ref{theorem3}, except that the channel state diagram has been changed and we must have:
\begin{equation*}
p(\tilde y\left| {{x_1}} \right.) = p(E\left| {{x_1}} \right.) = \frac{{p(E,{x_1})}}{{p({x_1})}} = \frac{{\sum\limits_{{y_2}} {p(E,{x_1},{y_2})} }}{{p({x_1})}}
\end{equation*}
\begin{equation*}
 = \frac{{\sum\limits_{{y_2}} {p({x_1})p({y_2}\left| {{x_1}} \right.)p(E\left| {{y_2}} \right.)} }}{{p({x_1})}}.
\end{equation*}
\end{IEEEproof}

\section{Numerical Results}
\label{numer}

\begin{figure}
\centering
\includegraphics[width=10cm]{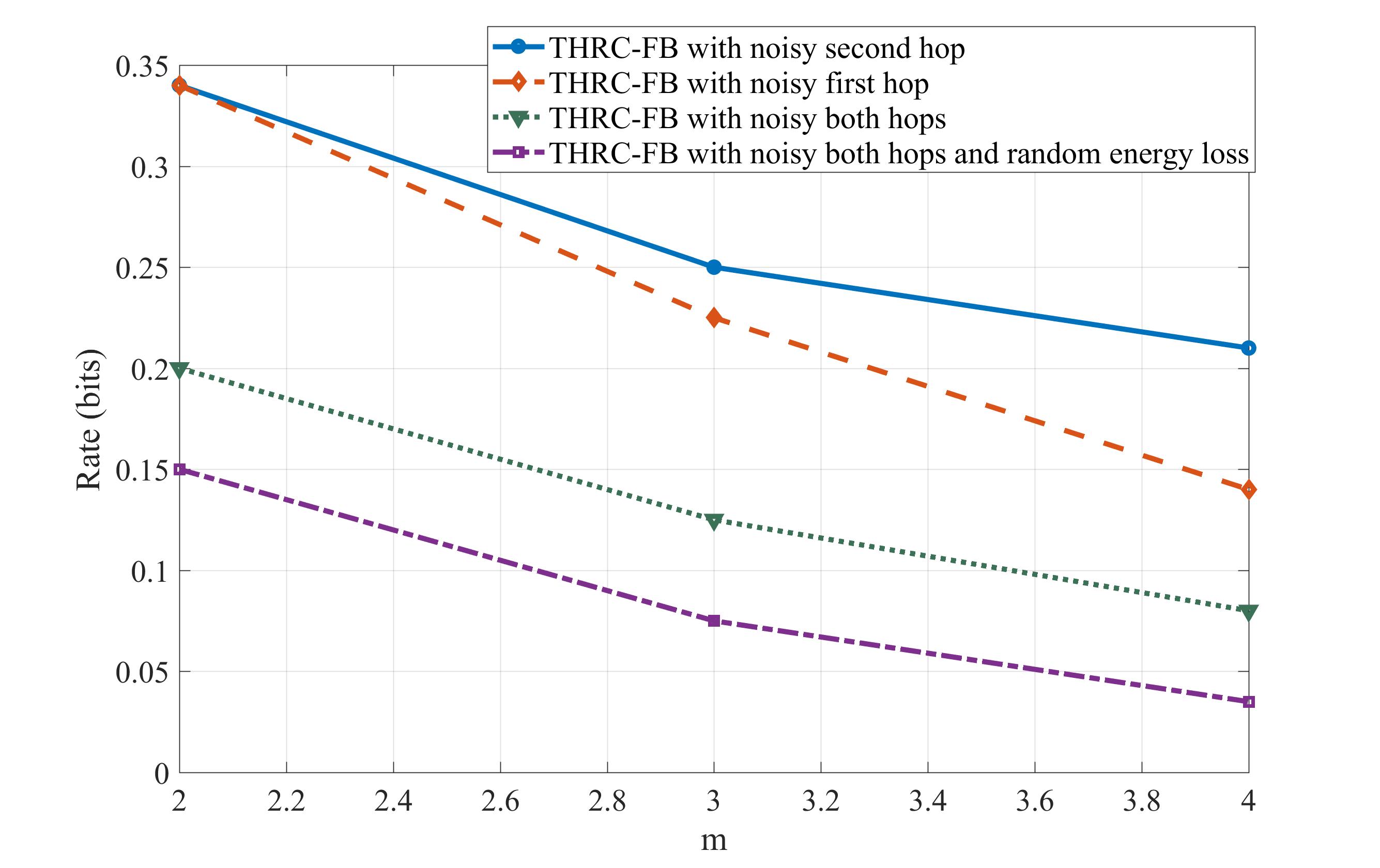}
\caption{Comparison of the achievable rates for different models with respect to $m$ when $U=m$.}
 \label{result1}
\end{figure}

\begin{figure}
\centering
\includegraphics[width=10cm]{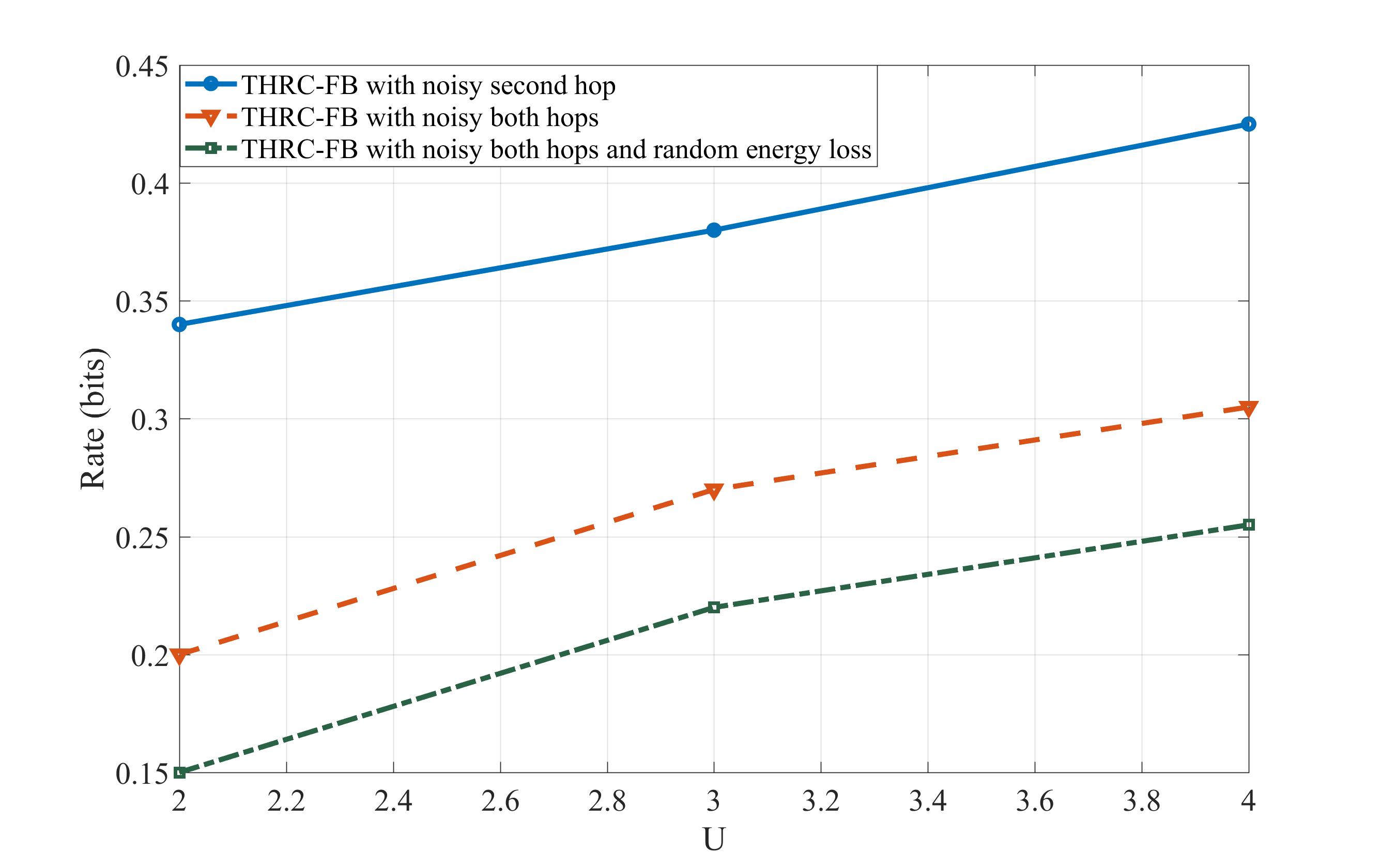}
\caption{Comparison of the achievable rates for different models with respect to $U$ when $m=2$.}
 \label{result2}
\end{figure}

Finally, a comparison of the achievable rates for our different models is illustrated in Figs \ref{result1} and \ref{result2}. In these figures, we consider a binary symmetric channel with parameter $p=0.1$ for the second hop and a binary symmetric channel with parameter $p=0.05$ for the first hop and for random energy loss, we conside $p(e=1|y_2=1)=0.99,$ $p(e=0|y_2=1)=0.1$. In addition, for timing approach, we consider $v=(U-Z)mod N+1,N=5$.  We can see that existence of noise in the first hop is more distruptive than the second hop.

\section{Discussions and Conclusions}
\label{section7}
We studied a binary two-hop channel with an energy harvesting relay with finite battery size, where the transmitter jointly transfers information and energy to the relay. Modeling the energy level at the relay's battery as channel states, we propose the achievability schemes for the channel with memory, where the main challenge was the unknown state at the receiver. We proposed achievable schemes for the cases with a noisy second hop, noisy first hop, and noisy both hops. We mention some of our future destinations as following:

\textbf{Upper bound}: Due to the channel memory, the problem of finding a tight outer bound for this system model cannot be tackled using standard inequalities used in converse proofs, thus, we have to find a tight upper bound for this problem.
%

\textbf{Battery size more than $m$ for timing approach}: If the battery size is more than $m$, the equivalent timing channel cannot be used because the timing noise, in this case, would have memory and the analysis would be more complicated, thus, we have to find a new scheme for this case.

\textbf{Finding capacity bounds for the case when there is a direct link between the transmitter and the receiver}: In our model, there was not a direct link between the transmitter and the receiver, so, we have to adapt our scheme to the case when there exists a direct link between the transmitter and the receiver.

\begin{appendices}

\section{}
\label{appendix 1}
It is known that a sufficient condition for the existence of the steady-state probabilities is that there exist a state $\hat{S}$ and a positive number $n$ such that beginning from any state we can reach the state $\hat{S}$ in $n$ steps \cite[Theorem 6.3.2.]{ash}. Now, we show these conditions hold by choosing $\hat{S}$ to be $\tilde{S}$ and $n$ to be the maximum distance between $\tilde{S}$ and any other states in the state diagram. Let $S^k$ be the set of states, which has distance $k$ from $\tilde{S}$ and $m$ be the maximum distance, i.e., $m=\max k$. One can reach $\tilde{S}$ from an element of $S^k$ in $l$ steps, where $l$ is an arbitrary integer number that $l\geq k$, Because after first arrival to $\tilde{S}$, we can stay there (due to the assumption of lemma). Therefore, beginning from any state we can reach $\tilde{S}$ in at least $m$ steps.

\section{}
\label{appendix 2}
Without loss of generality, we assume that $m=1$. Based on the packing Lemma \cite[Lemma 3.1]{elGamal}, if the sequence $X^n(1)$ is passed through a discrete memoryless channel $\prod\limits_{i = 1}^n {{p_{Y\left| X \right.}}(\left. {{y_i}} \right|{x_i})} $ and the sequence $Y^n$ is constructed and if $R < I(X;Y\left| U \right.)$, then we would have $P(\exists m \ne 1:({U^n},{X^n}(m),{Y^n}) \in A_\varepsilon ^{(n)}(p(u,x,y))) \to 0$. Now we consider a $p(y\left| x \right.)$  such  that $p(y = x\left| x \right.) = 1,p(y \ne x\left| x \right.) = 0$, so we obtain $Y^n=X^n(1)$. Now we determine $A_\varepsilon ^{(n)}(p(u,x,y))$. By the definition of the robust typicallity, we have:
\begin{equation*}
\scalebox{1}[1]{$A_\varepsilon ^{(n)}(p(u,x,y)) = \{ ({x^n},{u^n},{y^n})\left| {\left| {\pi (x,u,y\left| {{x^n},{u^n},{y^n}} \right.) - p(u,x,y)} \right|} \right. \le$}  
\end{equation*}
\begin{equation}
\varepsilon p(u,x,y),\forall u \in \mathcal{U},\forall x \in \mathcal{X},\forall y \in \mathcal{Y}\},
\label{Aeps}
\end{equation}
where ${\pi (x,u,y\left| {{x^n},{u^n},{y^n}} \right.)}$ is the percentage of repetition of $x,u,y$ in the sequence ${x^n},{u^n},{y^n}$. 

For the conditional distribution described above, we have $p\left( {u,x,y \ne x} \right) = 0,p\left( {u,x,y = x} \right) = p\left( {u,x} \right)$ $\times p\left( {y\left| x \right.} \right) = p\left( {u,x} \right)$, and thus:
\begin{equation}
\pi \left( {u,x,y \ne x\left| {{u^n},{x^n},{y^n}} \right.} \right) = 0,
\label{pi0}
\end{equation}
\begin{equation}
\pi \left( {u,x,y = x\left| {{u^n},{x^n},{y^n}} \right.} \right) = \pi \left( {u,x\left| {{u^n},{x^n}} \right.} \right).
\label{pi1}
\end{equation}

By inserting (\ref{pi0}) and (\ref{pi1}) in (\ref{Aeps}), we get:
\begin{equation}
\label{Aepsilon}
\scalebox{1}[1]{$A_\varepsilon ^{(n)}\left( {p\left( {u,x,y} \right)} \right) = \left\{ {\left( {{u^n},{x^n},{y^n}} \right)\left| {\left( {{u^n},{x^n}} \right) \in A_\varepsilon ^{(n)}\left( {p\left( {u,x} \right)} \right),} \right.{x^n} = {y^n}} \right\},$}
\end{equation}
which results in:
\begin{equation*}
p\left(\bigcup\limits_{m \ne 1} {\left\{ {({U^n},{X^n}(m),{Y^n}) \in A_\varepsilon ^{(n)}(p(u,x,y))} \right\}} \right)
\end{equation*}
\begin{equation*}
 = p\left(\bigcup\limits_{m \ne 1} {\left\{ {\{ ({U^n},{X^n}(m)) \in A_\varepsilon ^{(n)}(p(u,x))\}  \cap \{ {X^n}(m) = {X^n}(1)\} } \right\}} \right)
\end{equation*}
\begin{equation*}
 = p\left(\bigcup\limits_{m \ne 1} {\left\{ {\{ ({U^n},{X^n}(1)) \in A_\varepsilon ^{(n)}(p(u,x))\}  \cap \{ {X^n}(m) = {X^n}(1)\} } \right\}} \right).
\end{equation*}

So by packing lemma \cite{elGamal} we have:
\begin{equation*}
p\left(\bigcup\limits_{m \ne 1} {\left\{ {({U^n},{X^n}(m),{Y^n}) \in A_\varepsilon ^{(n)}(p(u,x,y))} \right\}} \right)
\end{equation*}
\begin{equation*}
 = p\left(\bigcup\limits_{m \ne 1} {\left\{ {\{ ({U^n},{X^n}(1)) \in A_\varepsilon ^{(n)}(p(u,x))\}  \cap \{ {X^n}(m) = {X^n}(1)\} } \right\}} \right) < \varepsilon '(n),
\end{equation*}
where $\mathop {\lim }\limits_{n \to \infty } \varepsilon '(n) = 0$. By complementing above equation we have:
\begin{equation}
 \scalebox{0.9}[1]{$p\left(\bigcap\limits_{m \ne 1} {\left\{ {\{ ({U^n},{X^n}(1)) \notin A_\varepsilon ^{(n)}(p(u,x))\}  \cup \{ {X^n}(m) \ne {X^n}(1)\} } \right\}} \right) > 1 - \varepsilon '(n),$}
\label{complement}
\end{equation}
and by distributivity of intersection over union, we have:
\begin{equation}
p\left(\left\{ {({U^n},{X^n}(1)) \notin A_\varepsilon ^{(n)}(p(u,x))} \right\} \cup \left\{ {\bigcap\limits_{m \ne 1} {\left\{ {{X^n}(m) \ne {X^n}(1)} \right\}} } \right\}\right) > 1 - \varepsilon '(n).
\end{equation}
we apply union bound to obtain:
\begin{equation*}
p\left(\left\{ {({U^n},{X^n}(1)) \notin A_\varepsilon ^{(n)}(p(u,x))} \right\} \cup \left\{ {\bigcap\limits_{m \ne 1} {\left\{ {{X^n}(m) \ne {X^n}(1)} \right\}} } \right\}\right)
\end{equation*}
\begin{equation}
 \scalebox{0.9}[1]{$\le p\left(\left\{ {({U^n},{X^n}(1)) \notin A_\varepsilon ^{(n)}(p(u,x))} \right\}\right) + p\left(\bigcap\limits_{m \ne 1} {\left\{ {{X^n}(m) \ne {X^n}(1)} \right\}} \right).$}
\label{pAeps}
\end{equation}

In addition, by the L.L.N, we have:
\begin{equation}
p\left(\left\{ {({U^n},{X^n}(1)) \notin A_\varepsilon ^{(n)}(p(u,x))} \right\}\right) < \varepsilon ''(n),
\label{jointAeps}
\end{equation}
where $\mathop {\lim }\limits_{n \to \infty } \varepsilon ''(n) = 0$. By combining (\ref{complement})-(\ref{jointAeps})  we obtain:
\begin{equation}
p\left(\bigcap\limits_{m \ne 1} {\left\{ {{X^n}(m) \ne {X^n}(1)} \right\}} \right) > 1 - \varepsilon '(n) - \varepsilon ''(n).
\end{equation}

So by packing lemma \cite[Lemma 3.1]{elGamal} and discussions given in the first paragraph of the proof and the fact $X=Y$, if we have:
\begin{equation*}
R < I\left( {X;Y\left| U \right.} \right)=I\left( {X;X\left| U \right.} \right) = H\left( {X\left| U \right.} \right),
\end{equation*}
then we can derive inequality (\ref{inequal}):
\begin{equation}
\label{inequal}
p\left(\bigcup\limits_{m \ne 1} {\left\{ {{X^n}(m) = {X^n}(1)} \right\}} \right) < \varepsilon '(n) +\varepsilon ''(n),
\end{equation} 
Where $\mathop {\lim }\limits_{n \to \infty } \varepsilon '(n)+\varepsilon ''(n) = 0$.

\section{}
\label{appendix 3}
Independence can be deduced from the codebook generation because the initial states and all codewords in each subcodebook are generated independently. To show that these sequences are regular Markov sources, we define a new Markov chain as $S_1=(U_1,U_2),S_2=(U_2,U_3),S_3=(U_3,U_4),\ldots$ (denoting $U_1,U_2,U_3,\ldots$ as the state sequence), as we described in section (II). By the assumption $m>1$ and the state diagram of the channel, it can be seen easily that ${X_2}_i$ is determined by $U_i,U_{i+1}$, so we have ${X_{2,i}}=f(S_i)$, where $f$ is a deterministic function. Thus, ${X_{2,i}}$ is a Markov source. Moreover, the assumptions of Lemma~\ref{lem:steady} are also satisfied by the new Markov chain $S_i$ and so the steady-state probabilities exist. This shows that ${X_{2,i}}$ is a regular Markov source.

\section{}
\label{appendix 4}

Consider the following relations:
\begin{equation*}
{q_1} = \frac{{{w_1}}}{{{j_1}}},{w_1} \in Z,{j_1} \in Z - \{ 0\}, 
\end{equation*}
\begin{equation*}
{q_2} = \frac{{{w_2}}}{{{j_2}}},{w_2} \in Z,{j_2} \in Z - \{ 0\} .
\end{equation*}

First of all, we introduce a class of finite-state channels and we prove regularity for them. Consider a finite set $S$ is the set of states, where $|S|=j_1j_2$. We consider the transition matrices $M_1$ and $M_2$ as the following:
\begin{equation}
{M_1} = {M_2} = \frac{1}{{{j_1}{j_2}}}{\left[ {\begin{array}{*{20}{c}}
1\\
1\\
 \vdots 
\end{array}\begin{array}{*{20}{c}}
1\\
1\\
{}
\end{array}\begin{array}{*{20}{c}}
 \cdots \\
{}\\
 \ddots 
\end{array}} \right]_{{j_1}{j_2} \times {j_1}{j_2}}},
\label{transition}
\end{equation}
where $M_1$ is the transition matrix when the input symbol is "0" and $M_2$ is the transition matrix when the input symbol is "1".
We can see that:
\begin{equation*}
{M_{i(1)}}{M_{i(2)}}...{M_{i(k)}} = M_1^k = M_2^k = {M_1} = {M_2},\forall i(1),...,i(k) \in \{ 1,2\} ,\forall k = \{ 1,2,...\}.
\end{equation*}
By \cite[Theorem 7.2.1.]{ash}, for a Markov chain with transition matrix $M_1$, steady-state probabilities exist because all columns of $M_1$ has nonzero elements. So a channel with the set of states $S$, the set of inputs $\Gamma=\{0,1\}$, the set of outputs $\Delta=\{0,1\}$ and transition matrices (\ref{transition}) is a regular finite-state channel. We call this class complete regular finite-state channels (because the state diagram of the channel for each input is a complete graph with equal values on each edge).

Now, we prove that the memoryless channel given by conditional p.m.f (\ref{conditional p.m.f}) is a special case of the class of complete regular finite-state channels. Consider $S_0$ as the initial state with p.m.f. $p(s_0)$, then we have:
\begin{equation*}
p({s_0},{s_1},{s_2},...,{s_n}) = p({s_0})p({s_1}\left| {{s_0}} \right.)p({s_2}\left| {{s_0},{s_1}} \right.)...p({s_n}\left| {{s_0},{s_1},{s_2},...} \right.),
\end{equation*} 
\begin{equation}
\left\{ {\begin{array}{*{20}{c}}
{p({s_1}\left| {{s_0}} \right.) = \frac{1}{{{j_1}{j_2}}}\left[ {\begin{array}{*{20}{c}}
1&1& \cdots &1
\end{array}} \right] = p({s_1})}\\
{p({s_2}\left| {{s_0},{s_1}} \right.) = p({s_2}\left| {{s_1}} \right.) = \frac{1}{{{j_1}{j_2}}}\left[ {\begin{array}{*{20}{c}}
1&1& \cdots &1
\end{array}} \right] = p({s_2})}\\
 \vdots \\
{p({s_n}\left| {{s_0},{s_1},{s_2},...} \right.) = p({s_n}\left| {{s_{n - 1}}} \right.) = \frac{1}{{{j_1}{j_2}}}\left[ {\begin{array}{*{20}{c}}
1&1& \cdots &1
\end{array}} \right] = p({s_n})}
\end{array}} \right.,
\label{markov}
\end{equation}
where equations (\ref{markov}) can be easily derived by the special form of transition matrices $M_1$ and $M_2$. Then, we have:
\begin{equation*}
p({s_0},{s_1},{s_2},...,{s_n}) = p({s_0})p({s_1})p({s_2})...p({s_n}),
\end{equation*}
\begin{equation}
p({s_1},{s_2},...,{s_n}) = p({s_1})p({s_2})...p({s_n}).
\label{i.i.d}
\end{equation}
By equations (\ref{markov}) and (\ref{i.i.d}), we can see that the sequence $s_1,...,s_n$ is i.i.d. By the definition of finite state channels, the output of the channel in the $i$-th channel use can be written as follows:
\begin{equation}
y_i=g(x_i,s_i),
\end{equation}
where $g$ is a deterministic function. So, we can write:
\begin{equation*}
p({y^n}\left| {{x^n}} \right.) = p(g({x_1},{s_1}),g({x_2},{s_2}),....\left| {{x^n}} \right.)
\end{equation*}
\begin{equation}
 = p(g({x_1},{s_1})\left| {{x^n}} \right.)p(g({x_2},{s_2})\left| {{x^n}} \right.)... = \prod\limits_{i = 1}^n {p({y_i}\left| {{x_i}} \right.)}.
\label{i.i.d2}
\end{equation}
Equation (\ref{i.i.d2}) can be concluded because the sequence $s_1,...,s_n$ is i.i.d. Until this step, we have proved that the class of complete finite state channels are memoryless. In the last step, we choose the function $g$ such that a complete finite state channel's conditional p.m.f becomes in the form of (\ref{conditional p.m.f}). Consider two stes $S_1$ and $S_2$ such that ${S_1},{S_2} \subseteq S,\left| {{S_1}} \right| = {w_1}{j_2},\left| {{S_2}} \right| = {w_2}{j_1}$. Then, we consider $g$ in this way:

\begin{equation}
\left\{ {\begin{array}{*{20}{c}}
{g(x = 0,s) = 0,s \in {S_1}}\\
{g(x = 0,s) = 1,s \in S - {S_1}}\\
{g(x = 1,s) = 1,s \in {S_2}}\\
{g(x = 1,s) = 0,s \in S - {S_2}}
\end{array}} \right.,
\end{equation}
so we can easily see that:
\begin{equation}
p(y = 0\left| {x = 0} \right.) = p(s \in {S_1}) = \frac{{\left| {{S_1}} \right|}}{{\left| S \right|}} = \frac{{{w_1}{j_2}}}{{{j_1}{j_2}}} = {q_1},
\end{equation}
\begin{equation}
p(y = 1\left| {x = 1} \right.) = p(s \in {S_2}) = \frac{{\left| {{S_2}} \right|}}{{\left| S \right|}} = \frac{{{w_2}{j_1}}}{{{j_1}{j_2}}} = {q_2}.
\end{equation}

So the memoryless channel with conditional p.m.f (\ref{conditional p.m.f}) with rational values of $q_1$ and $q_2$, is a regular finite state channel and by \cite[Theorem 7.2.1.]{ash}, the sequences $Y^n$ and $(X^n,Y^n)$ are regular Markov sources and obey A.E.P, so equations (\ref{a.e.p.rat}) and (\ref{a.e.p.rat2}) has been proved.

\section{}
\label{appendix 5}

We calculate the conditional p.m.f of  $Y$ and put it equal to the conditional p.m.f (\ref{conditional p.m.f}):
\begin{equation}
p(y = 0\left| {x = 0} \right.) = q_1^k\tilde q_1^k + (1 - q_1^k)(1 - \tilde q_2^k) = {q_1},
\label{equal1}
\end{equation}
\begin{equation}
p(y = 1\left| {x = 1} \right.) = q_2^k\tilde q_2^k + (1 - q_2^k)(1 - \tilde q_1^k) = {q_2}.
\label{equal2}
\end{equation}
The reason of the assuming $q_1^k+q_2^k\ne1, \forall k \in \{1,2,...\}$ is solvability of equations (\ref{equal1}) and (\ref{equal2}). If we solve equations (\ref{equal1}) and (\ref{equal2}), we would have:
\begin{equation}
\tilde q_1^k = \frac{{q_2^k{q_1} - q_1^k{q_2}}}{{q_1^k + q_2^k - 1}} + \frac{{q_1^k + {q_2} - 1}}{{q_1^k + q_2^k - 1}},
\label{q_tilde1}
\end{equation}
\begin{equation}
\tilde q_2^k =  - \frac{{q_2^k{q_1} - q_1^k{q_2}}}{{q_1^k + q_2^k - 1}} + \frac{{q_2^k + {q_1} - 1}}{{q_1^k + q_2^k - 1}}.
\label{q_tilde2}
\end{equation}
It can be easily seen that:
\begin{equation*}
\mathop {\lim }\limits_{k \to \infty } \tilde q_1^k = \mathop {\lim }\limits_{k \to \infty }\tilde q_2^k  = 1,
\end{equation*}
so, for large enough $k$, we have $\tilde q_1^k>0$ and $\tilde q_2^k>0$ .

First, we consider $q_2 + {q_1} - 1>1$. We set $q_1^k=q_1+\alpha_k$ and $q_2^k=q_2+\beta_k$, where $\mathop {\lim }\limits_{k \to \infty } \alpha_k=\mathop {\lim }\limits_{k \to \infty }\beta_k=0$. So, $\alpha_k$ and $\beta_k$ must be chosen in a way such that $\tilde q_1^k<1$ and $\tilde q_2^k<1$. By applying these coditions to (\ref{q_tilde1}) and (\ref{q_tilde2}), for large enough $k$, we have $q_1^k+q_2^k-1>1$, so we conclude the coditions below:
\begin{equation}
q_2^k{q_1} - q_1^k{q_2} + q_1^k + {q_2} - 1 < q_1^k + q_2^k - 1,
\end{equation}
\begin{equation}
 - q_2^k{q_1} + q_1^k{q_2} + q_2^k + {q_1} - 1 < q_1^k + q_2^k - 1,
\end{equation}
for which we have:
\begin{equation}
{\beta _k}( {q_1}-1) < {\alpha _k}{q_2},
\label{cond1}
\end{equation}
\begin{equation}
{\alpha _k}({q_2}-1) < {\beta _k}{q_1}.
\label{cond2}
\end{equation}
We can see that, if we choose $\alpha_k,\beta_k>0$, conditions (\ref{cond1}) and (\ref{cond2}) will be satisfied (it means that $q_1^k$ and $q_2^k$ must tend to $q_1$ and $q_2$ from the upside).
When $q_2 + {q_1} - 1<1$, by similar arguments, we can conclude that $\alpha_k,\beta_k<0$ and $q_1^k$ and $q_2^k$ must tend to $q_1$ and $q_2$ from the downside. So, The memoryless channel between $\tilde Y_{[k]}$ and $Y$ with parameters $0\le \tilde q_1^k \le 1$ and $0\le \tilde q_2^k \le 1$ exists for large enough $k$.

\section{}
\label{appendix 6}
We have:
\begin{equation}
p\left( {\left| {\frac{1}{n}\log \left( {\frac{{p(\tilde Y_{[k]}^n,{Y^n})}}{{p({Y^n})}}} \right)} \right| > \varepsilon } \right) \le \frac{1}{{n\varepsilon }}E\left\{ {\left| {\log \left( {\frac{{p(\tilde Y_{[k]}^n,{Y^n})}}{{p({Y^n})}}} \right)} \right|} \right\}
\label{lemma6_1}
\end{equation}
\begin{equation}
 \le \frac{1}{{n\varepsilon }}\left\{ {2\frac{{\log e}}{e}E\left\{ {\frac{{p(\tilde Y_{[k]}^n,{Y^n})}}{{p({Y^n})}}} \right\} - E\left\{ {\log \left( {\frac{{p(\tilde Y_{[k]}^n,{Y^n})}}{{p({Y^n})}}} \right)} \right\}} \right\},
\label{lemma6_2}
\end{equation}
where, (\ref{lemma6_1}) is written by Markov's inequality and ($\ref{lemma6_2}$) is written by (\ref{abs1}) and (\ref{abs2}). The first term in (\ref{lemma6_2}) can be bounded as:
\begin{equation*}
E\left\{ {\frac{{p(\tilde Y_{[k]}^n,{Y^n})}}{{p({Y^n})}}} \right\} = \sum\limits_{\tilde y_{[k]}^n,{y^n}} {\frac{{p(\tilde y_{[k]}^n,{y^n})}}{{p({y^n})}}p({y^n})p(\tilde y_{[k]}^n\left| {{y^n}} \right.)}  \le \sum\limits_{\tilde y_{[k]}^n,{y^n}} {p(\tilde y_{[k]}^n,{y^n})}  = 1,
\end{equation*}
because $p(\tilde y_{[k]}^n\left| {{y^n}} \right.)  \le 1$. For the second term we have:
\begin{equation*}
 - E\left\{ {\log \left( {\frac{{p(\tilde Y_{[k]}^n,{Y^n})}}{{p({Y^n})}}} \right)} \right\} =  - E\left\{ {\log \left( {p(\tilde Y_{[k]}^n\left| {{Y^n}} \right.)} \right)} \right\} = H(\tilde Y_{[k]}^n\left| {{Y^n}} \right.) \le n\mathop {\max }\limits_i H({{\tilde Y}_{[k],i}}\left| {{Y_i}} \right.).
\end{equation*}
So, we have:
\begin{equation}
p\left( {\left| {\frac{1}{n}\log \left( {\frac{{p(\tilde Y_{[k]}^n,{Y^n})}}{{p({Y^n})}}} \right)} \right| > \varepsilon } \right) \le 2\frac{{\log e}}{{n\varepsilon e}} +\frac{{1}}{{\varepsilon }} \mathop {\max }\limits_i H({{\tilde Y}_{[k],i}}\left| {{Y_i}} \right.).
\label{probconv1}
\end{equation}
By tending $n$ and $k$ to infinity, we have:
\begin{equation}
\mathop {\lim }\limits_{n,k \to \infty }  - \frac{1}{n}\log\left( {\frac{{p(\tilde Y_{[k]}^n,{Y^n})}}{{p({Y^n})}}} \right)\xrightarrow{prob.}0.
\label{a.e.p.rat3}
\end{equation}

\section{}
\label{appendix 7}
By the argument similar to Lemma \ref{lemma6} we have:
\begin{equation}
p\left( {\left| {\frac{1}{n}\log \left( {\frac{{p(\tilde Y_{[k]}^n,{Y^n})}}{{p(\tilde Y_{[k]}^n)}}} \right)} \right| > \varepsilon } \right) \le \frac{1}{{n\varepsilon }}E\left\{ {\left| {\log \left( {\frac{{p(\tilde Y_{[k]}^n,{Y^n})}}{{p(\tilde Y_{[k]}^n)}}} \right)} \right|} \right\}
\label{lemma6_3}
\end{equation}
\begin{equation}
 \le \frac{1}{{n\varepsilon }}\left\{ {2\frac{{\log e}}{e}E\left\{ {\frac{{p(\tilde Y_{[k]}^n,{Y^n})}}{{p(\tilde Y_{[k]}^n)}}} \right\} - E\left\{ {\log \left( {\frac{{p(\tilde Y_{[k]}^n,{Y^n})}}{{p(\tilde Y_{[k]}^n)}}} \right)} \right\}} \right\},
\label{lemma6_4}
\end{equation}
where, (\ref{lemma6_3}) is written by Markov's inequality and ($\ref{lemma6_4}$) is written by (\ref{abs1}) and (\ref{abs2}). The first term in (\ref{lemma6_4}) can be bounded as:
\begin{equation*}
E\left\{ {\frac{{p(\tilde Y_{[k]}^n,{Y^n})}}{{p(\tilde Y_{[k]}^n)}}} \right\} = \sum\limits_{\tilde y_{[k]}^n,{y^n}} {\frac{{p(\tilde y_{[k]}^n,{y^n})}}{{p(\tilde y_{[k]}^n)}}p(\tilde y_{[k]}^n)p({y^n}\left| {\tilde y_{[k]}^n} \right.)}  \le \sum\limits_{\tilde y_{[k]}^n,{y^n}} {p(\tilde y_{[k]}^n,{y^n})}  = 1,
\end{equation*}
because $p(y^n\left| {{\tilde y_{[k]}^n}} \right.)  \le 1$. For the second term we have:
\begin{equation*}
 - E\left\{ {\log \left( {\frac{{p(\tilde Y_{[k]}^n,{Y^n})}}{{p(\tilde Y_{[k]}^n)}}} \right)} \right\} =  - E\left\{ {\log \left( {p({Y^n}\left| {\tilde Y_{[k]}^n} \right.)} \right)} \right\} = H({Y^n}\left| {\tilde Y_{[k]}^n} \right.) \le n\mathop {\max }\limits_i H({Y_i}\left| {{{\tilde Y}_{[k],i}}} \right.).
\end{equation*}
So, we have:
\begin{equation}
p\left( {\left| {\frac{1}{n}\log \left( {\frac{{p(\tilde Y_{[k]}^n,{Y^n})}}{{p(\tilde Y_{[k]}^n)}}} \right)} \right| > \varepsilon } \right) \le 2\frac{{\log e}}{{n\varepsilon e}} + \frac{1}{\varepsilon }\mathop {\max }\limits_i H({Y_i}\left| {{{\tilde Y}_{[k],i}}} \right.).
\label{probconv2}
\end{equation}
By tending $n$ and $k$ to infinity, we have:
\begin{equation}
\mathop {\lim }\limits_{n,k \to \infty }  - \frac{1}{n}\log\left( {\frac{{p(\tilde Y_{[k]}^n,{Y^n})}}{{p(\tilde Y_{[k]}^n)}}} \right) \xrightarrow{prob.}0.
\label{a.e.p.rat4}
\end{equation}

\section{}
\label{appendix 8}

First of all, we show that $H\{\underline{Y}\}$ is close to $H\left\{ {\underline{{\tilde Y}_{[k]}}} \right\}$ for large enough $k$. By properties of entropy, we can write:
\begin{equation*}
\frac{1}{n}H(\tilde Y_{[k]}^n) + \frac{1}{n}H({Y^n}\left| {\tilde Y_{[k]}^n} \right.) = \frac{1}{n}H({Y^n}) + \frac{1}{n}H(\tilde Y_{[k]}^n\left| {{Y^n}} \right.),
\end{equation*}
\begin{equation*}
\left| {\frac{1}{n}H({Y^n}) - \frac{1}{n}H(\tilde Y_{[k]}^n)} \right| = \left| {\frac{1}{n}H({Y^n}\left| {\tilde Y_{[k]}^n} \right.) - \frac{1}{n}H(\tilde Y_{[k]}^n\left| {{Y^n}} \right.)} \right|
\end{equation*}
\begin{equation}
 \le \frac{1}{n}H({Y^n}\left| {\tilde Y_{[k]}^n} \right.) + \frac{1}{n}H(\tilde Y_{[k]}^n\left| {{Y^n}} \right.) \le \mathop {\max }\limits_i H({Y_i}\left| {{{\tilde Y}_{[k],i}}} \right.) + \mathop {\max }\limits_i H({{\tilde Y}_{[k],i}}\left| {{Y_i}} \right.),
\label{probconv3}
\end{equation}
so, for sufficiently large $n$ and $k$, $H\{\underline{Y}\}$ is close to $\frac{1}{n}H({Y^n})$, $H\left\{ {\underline{{\tilde Y}_{[k]}}} \right\}$ is close to $\frac{1}{n}H(\tilde Y_{[k]}^n)$  and $\frac{1}{n}H(\tilde Y_{[k]}^n)$ is close to $\frac{1}{n}H({Y^n})$. Now, to prove that $ - \frac{1}{n}\log (p({Y^n}))$ converges in probability to $H\{\underline{Y}\}$, we can write:
\begin{equation*}
p\left(\left| { - \frac{1}{n}\log (p({Y^n})) - H\left\{ \underline{Y} \right\}} \right| > \varepsilon \right) \le 
\end{equation*} 
\begin{equation*}
p\left(\left\{ {\left| {\frac{1}{n}\log \left( {\frac{{p(\tilde Y_{[k]}^n,{Y^n})}}{{p({Y^n})}}} \right)} \right| > \frac{\varepsilon }{4}} \right\} \cup \left\{ {\left| {\frac{1}{n}\log \left( {\frac{{p(\tilde Y_{[k]}^n,{Y^n})}}{{p(\tilde Y_{[k]}^n)}}} \right)} \right| > \frac{\varepsilon }{4}} \right\}\right.
\end{equation*}
\begin{equation}
 \left. \cup \left\{ {\left| { - \frac{1}{n}\log (p({\tilde Y_{[k]}^n})) - H\left\{ {\underline{{\tilde Y}_{[k]}}} \right\}} \right| > \frac{\varepsilon }{4}} \right\} \cup \left\{ {\left| {H\left\{\underline{Y} \right\} - H\left\{ {\underline{{\tilde Y}_{[k]}}} \right\}} \right|} \right\} > \frac{\varepsilon }{4}\right),
\end{equation}
because, if all of four events in the above union  does not occur, then we have $\left| { - \frac{1}{n}\log (p({Y^n})) - H\left\{ Y \right\}} \right| \le \varepsilon $. By the union bound, we can write:
\begin{equation*}
p\left(\left| { - \frac{1}{n}\log (p({Y^n})) - H\left\{ \underline{Y} \right\}} \right| > \varepsilon \right) \le
\end{equation*}
\begin{equation*}
p\left( {\left| {\frac{1}{n}\log \left( {\frac{{p(\tilde Y_{[k]}^n,{Y^n})}}{{p({Y^n})}}} \right)} \right| > \frac{\varepsilon }{4}} \right) + p\left( {\left| {\frac{1}{n}\log \left( {\frac{{p(\tilde Y_{[k]}^n,{Y^n})}}{{p(\tilde Y_{[k]}^n)}}} \right)} \right| > \frac{\varepsilon }{4}} \right)
\end{equation*}
\begin{equation}
+p\left( {\left| { - \frac{1}{n}\log (p({\tilde Y_{[k]}^n})) - H\left\{ \underline{\tilde Y_{[k]}} \right\}} \right| > \frac{\varepsilon }{4}} \right) + p\left( {\left| {H\left\{ \underline{Y} \right\} - H\left\{ {\underline{{\tilde Y}_{[k]}}} \right\}} \right|}  > \frac{\varepsilon }{4}\right).
\end{equation}
By equations (\ref{probconv1}), (\ref{probconv2}), (\ref{probconv3}) and Lemma \ref{lemma4}, we can choose  large enough numbers $N(\varepsilon,\delta)$ and $K(\varepsilon,\delta)$, such that for $\forall n>N(\varepsilon,\delta)$ and $k=K(\varepsilon,\delta)$, equations (\ref{eqconv1})-(\ref{eqconv2}) are satisfied:
\begin{equation}
p\left( {\left| {\frac{1}{n}\log \left( {\frac{{p(\tilde Y_{[k]}^n,{Y^n})}}{{p({Y^n})}}} \right)} \right| > \frac{\varepsilon }{4}} \right) < \frac{\delta }{3},
\label{eqconv1}
\end{equation}
\begin{equation}
p\left( {\left| {\frac{1}{n}\log \left( {\frac{{p(\tilde Y_{[k]}^n,{Y^n})}}{{p(\tilde Y_{[k]}^n)}}} \right)} \right| > \frac{\varepsilon }{4}} \right) < \frac{\delta }{3},
\end{equation}
\begin{equation}
p\left( {\left| { - \frac{1}{n}\log (p({\tilde Y_{[k]}^n})) - H\left\{\underline{\tilde Y_{[k]}} \right\}} \right| > \frac{\varepsilon }{4}} \right) < \frac{\delta }{3},
\end{equation}
\begin{equation}
\left| {H\left\{ \underline{Y} \right\} - H\left\{ {\underline{{\tilde Y}_{[k]}}} \right\}} \right| <\frac{\varepsilon }{4}.
\label{eqconv2}
\end{equation}
So, for $\forall n>N(\varepsilon,\delta)$, we have:
\begin{equation*}
p\left(\left| { - \frac{1}{n}\log (p({Y^n})) - H\left\{ \underline{Y} \right\}} \right| > \varepsilon \right) <\delta,
\end{equation*}
and we can see:
\begin{equation*}
\mathop {\lim }\limits_{n \to \infty }  - \frac{1}{n}\log (p({Y^n}))\xrightarrow{prob.} H\{\underline{Y}\}.
\end{equation*}

\section{}
\label{appendix 9}

By similar arguments introduced in Lemma \ref{lemma6} and \ref{lemma7}, we have:
\begin{equation*}
p\left( {\left| {\frac{1}{n}\log \left( {\frac{{p({X^n},\tilde Y_{[k]}^n,{Y^n})}}{{p({X^n},{Y^n})}}} \right)} \right| > \varepsilon } \right) \le \frac{1}{{n\varepsilon }}E\left\{ {\left| {\log \left( {\frac{{p({X^n},\tilde Y_{[k]}^n,{Y^n})}}{{p({X^n},{Y^n})}}} \right)} \right|} \right\}
\end{equation*}
\begin{equation}
 \le \frac{1}{{n\varepsilon }}\left\{ {2\frac{{\log e}}{e}E\left\{ {\frac{{p({X^n},\tilde Y_{[k]}^n,{Y^n})}}{{p({X^n},{Y^n})}}} \right\} - E\left\{ {\log \left( {\frac{{p({X^n},\tilde Y_{[k]}^n,{Y^n})}}{{p({X^n},{Y^n})}}} \right)} \right\}} \right\},
\end{equation}
\begin{equation*}
E\left\{ {\frac{{p({X^n},\tilde Y_{[k]}^n,{Y^n})}}{{p({X^n},{Y^n})}}} \right\} = \sum\limits_{{x^n},\tilde y_{[k]}^n,{y^n}} {\frac{{p({x^n},\tilde y_{[k]}^n,{y^n})}}{{p({x^n},{y^n})}}p({x^n},{y^n})p(\tilde y_{[k]}^n\left| {{x^n},{y^n}} \right.)}  \le \sum\limits_{{x^n},\tilde y_{[k]}^n,{y^n}} {p({x^n},\tilde y_{[k]}^n,{y^n})}  = 1,
\end{equation*}
\begin{equation*}
\scalebox{.99}[1]{$- E\left\{ {\log \left( {\frac{{p({X^n},\tilde Y_{[k]}^n,{Y^n})}}{{p({X^n},{Y^n})}}} \right)} \right\} =  - E\left\{ {\log \left( {p(\tilde Y_{[k]}^n\left| {{X^n},{Y^n}} \right.)} \right)} \right\} = H(\tilde Y_{[k]}^n\left| {{X^n},{Y^n}} \right.) \le n\mathop {\max }\limits_i H({{\tilde Y}_{[k],i}}\left| {{Y_i}} \right.),$}
\end{equation*}
so, we will have:
\begin{equation}
p\left( {\left| {\frac{1}{n}\log \left( {\frac{{p({X^n},\tilde Y_{[k]}^n,{Y^n})}}{{p({X^n},{Y^n})}}} \right)} \right| > \varepsilon } \right) \le 2\frac{{\log e}}{{n\varepsilon e}} + \frac{1}{\varepsilon }\mathop {\max }\limits_i H({{\tilde Y}_{[k],i}}\left| {{Y_i}} \right.).
\label{probconvergence1}
\end{equation}
By tending $n$ and $k$ to infinity, we can write:
\begin{equation}
{ - \frac{1}{n}\log \left( {\frac{{p({X^n},\tilde Y_{[k]}^n,{Y^n})}}{{p({X^n},{Y^n})}}} \right)}\xrightarrow{prob.}0.
\label{a.e.p.rat3}
\end{equation}

\section{}
\label{appendix 10}

by similar arguments introduced in Lemma \ref{lemma6}, \ref{lemma7} and \ref{lemma9}, we have:
\begin{equation*}
p\left( {\left| {\frac{1}{n}\log \left( {\frac{{p({X^n},\tilde Y_{[k]}^n,{Y^n})}}{{p({X^n},\tilde Y_{[k]}^n)}}} \right)} \right| > \varepsilon } \right) \le \frac{1}{{n\varepsilon }}E\left\{ {\left| {\log \left( {\frac{{p({X^n},\tilde Y_{[k]}^n,{Y^n})}}{{p({X^n},\tilde Y_{[k]}^n)}}} \right)} \right|} \right\}
\end{equation*}
\begin{equation}
 \le \frac{1}{{n\varepsilon }}\left\{ {2\frac{{\log e}}{e}E\left\{ {\frac{{p({X^n},\tilde Y_{[k]}^n,{Y^n})}}{{p({X^n},\tilde Y_{[k]}^n)}}} \right\} - E\left\{ {\log \left( {\frac{{p({X^n},\tilde Y_{[k]}^n,{Y^n})}}{{p({X^n},\tilde Y_{[k]}^n)}}} \right)} \right\}} \right\},
\end{equation}
\begin{equation*}
E\left\{ {\frac{{p({X^n},\tilde Y_{[k]}^n,{Y^n})}}{{p({X^n},\tilde Y_{[k]}^n)}}} \right\} = \sum\limits_{{x^n},\tilde y_{[k]}^n,{y^n}} {\frac{{p({x^n},\tilde y_{[k]}^n,{y^n})}}{{p({x^n},\tilde y_{[k]}^n)}}p({x^n},\tilde y_{[k]}^n)p({y^n}\left| {{x^n},\tilde y_{[k]}^n} \right.)}  \le \sum\limits_{{x^n},\tilde y_{[k]}^n,{y^n}} {p({x^n},\tilde y_{[k]}^n,{y^n})}  = 1,
\end{equation*}
\begin{equation*}
\scalebox{.99}[1]{$ - E\left\{ {\log \left( {\frac{{p({X^n},\tilde Y_{[k]}^n,{Y^n})}}{{p({X^n},\tilde Y_{[k]}^n)}}} \right)} \right\} =  - E\left\{ {\log \left( {p({Y^n}\left| {{X^n},\tilde Y_{[k]}^n} \right.)} \right)} \right\} = H({Y^n}\left| {{X^n},\tilde Y_{[k]}^n} \right.) \le n\mathop {\max }\limits_i H({Y_i}\left| {{{\tilde Y}_{[k],i}}} \right.).$}
\end{equation*}
So, we will have:
\begin{equation}
p\left( {\left| {\frac{1}{n}\log \left( {\frac{{p({X^n},\tilde Y_{[k]}^n,{Y^n})}}{{p({X^n},\tilde Y_{[k]}^n)}}} \right)} \right| > \varepsilon } \right) \le 2\frac{{\log e}}{{n\varepsilon e}} + \frac{1}{\varepsilon }\mathop {\max }\limits_i H({Y_i}\left| {{{\tilde Y}_{[k],i}}} \right.).
\label{probconvergence2}
\end{equation}
By tending $n$ and $k$ to infinity, we can write:
\begin{equation}
{ - \frac{1}{n}\log \left( {\frac{{p({X^n},\tilde Y_{[k]}^n,{Y^n})}}{{p({X^n},\tilde Y_{[k]}^n)}}} \right)}\xrightarrow{prob.}0.
\label{a.e.p.rat3}
\end{equation}

\section{}
\label{appendix 11}

Similar to the proof of  Lemma \ref{lemma8}, we have:
\begin{equation*}
\left| {\frac{1}{n}H({X^n},{Y^n}) - \frac{1}{n}H({X^n},\tilde Y_{[k]}^n)} \right| \le 
\end{equation*}
\begin{equation}
\frac{1}{n}H(\tilde Y_{[k]}^n\left| {{X^n},{Y^n}} \right.) + \frac{1}{n}H({Y^n}\left| {{X^n},\tilde Y_{[k]}^n} \right.) \le \mathop {\max }\limits_i H({Y_i}\left| {{{\tilde Y}_{[k],i}}} \right.) + \mathop {\max }\limits_i H({{\tilde Y}_{[k],i}}\left| {{Y_i}} \right.),
\label{probconvergence3}
\end{equation}
thus, by choosing sufficiently large $n$ and $k$, we can see that $H\{\underline{X},\underline{Y}\}$ is close to $H\left\{\underline{X},\underline{Y_{[k]}}\right\}$. By similar arguments given in Lemma \ref{lemma8}, we have:
\begin{equation*}
p\left(\left| { - \frac{1}{n}\log (p({X^n,Y^n})) - H\left\{\underline{X}, \underline{Y} \right\}} \right| > \varepsilon \right) \le
\end{equation*}
\begin{equation*}
p\left( {\left| {\frac{1}{n}\log \left( {\frac{{p(X^n,\tilde Y_{[k]}^n,{Y^n})}}{{p({X^n,Y^n})}}} \right)} \right| > \frac{\varepsilon }{4}} \right) + p\left( {\left| {\frac{1}{n}\log \left( {\frac{{p(X^n,\tilde Y_{[k]}^n,{Y^n})}}{{p(X^n,\tilde Y_{[k]}^n)}}} \right)} \right| > \frac{\varepsilon }{4}} \right)
\end{equation*}
\begin{equation}
p\left( {\left| { - \frac{1}{n}\log (p({X^n,\tilde Y_{[k]}^n})) - H\left\{\underline{X}, \underline{\tilde Y_{[k]}} \right\}} \right| > \frac{\varepsilon }{4}} \right) + p\left( {\left| {H\left\{\underline{X}, \underline{Y} \right\} - H\left\{\underline{X}, {\underline{{\tilde Y}_{[k]}}} \right\}} \right|}  > \frac{\varepsilon }{4}\right).
\end{equation}
By equations (\ref{probconvergence1}), (\ref{probconvergence2}), (\ref{probconvergence3}) and Lemma \ref{lemma4}, we can choose  large enough numbers $N'(\varepsilon,\delta)$ and $K'(\varepsilon,\delta)$, such that for $\forall n>N'(\varepsilon,\delta)$ and $k=K'(\varepsilon,\delta)$, equations (\ref{eqconvergence1})-(\ref{eqconvergence2}) are satisfied:
\begin{equation}
p\left( {\left| {\frac{1}{n}\log \left( {\frac{{p({X^n},\tilde Y_{[k]}^n,{Y^n})}}{{p({X^n},{Y^n})}}} \right)} \right| > \frac{\varepsilon }{4}} \right)< \frac{\delta}{3},
\label{eqconvergence1}
\end{equation}
\begin{equation}
p\left( {\left| {\frac{1}{n}\log \left( {\frac{{p({X^n},\tilde Y_{[k]}^n,{Y^n})}}{{p({X^n},\tilde Y_{[k]}^n)}}} \right)} \right| > \frac{\varepsilon }{4}} \right)< \frac{\delta}{3},
\end{equation}
\begin{equation}
p\left( {\left| { - \frac{1}{n}\log (p({X^n,\tilde Y_{[k]}^n})) - H\left\{\underline{X}, \underline{\tilde Y_{[k]}} \right\}} \right| > \frac{\varepsilon }{4}} \right) < \frac{\delta}{3},
\end{equation}
\begin{equation}
 {\left| {H\left\{\underline{X}, \underline{Y} \right\} - H\left\{\underline{X}, {\underline{{\tilde Y}_{[k]}}} \right\}} \right|} < \frac{\varepsilon }{4}.
\label{eqconvergence2}
\end{equation}

So, for $\forall n>N'(\varepsilon,\delta)$ we have:
\begin{equation*}
p\left(\left| { - \frac{1}{n}\log (p({X^n,Y^n})) - H\left\{\underline{X}, \underline{Y} \right\}} \right| > \varepsilon \right) <\delta,
\end{equation*}
and we can see:
\begin{equation*}
\mathop {\lim }\limits_{n \to \infty }  - \frac{1}{n}\log (p({X^n,Y^n}))\xrightarrow{prob.} H\{\underline{X},\underline{Y}\}.
\end{equation*}

\section{}
\label{appendix 12}

Without loss of generality, we assume that $m=1$. Based on the covering Lemma \cite[Lemma 3.3]{elGamal}, if the sequence $X^n(1)$ is passed through a discrete memoryless channel $\prod\limits_{i = 1}^n {{p_{Y\left| X \right.}}(\left. {{y_i}} \right|{x_i})} $ and the sequence $Y^n$ is constructed and if $R > I(X;Y\left| U \right.)$, then we would have $p(\exists m \ne 1:({U^n},{X^n}(m),{Y^n}) \in A_\varepsilon ^{(n)}(p(u,x,y))) \to 1$. Now, we consider a $p(y\left| x \right.)$  such  that $p(y = x\left| x \right.) = 1,p(y \ne x\left| x \right.) = 0$, so we obtain $Y^n=X^n(1)$.  By the same argument given in Lemma \ref{lemmaaepsilon}, we can easily see that if $R>H(X|U)$, then we have:
\begin{equation}
p\left(\left\{ {({U^n},{X^n}(1)) \notin A_\varepsilon ^{(n)}(p(u,x))} \right\} \cup \left\{ {\bigcap\limits_{m \ne 1} {\left\{ {{X^n}(m) \ne {X^n}(1)} \right\}} } \right\}\right) < \varepsilon '(n),
\end{equation}
where $\mathop {\lim }\limits_{n \to \infty } \varepsilon '(n) = 0$, and by the next inequality, the proof is complete.
\begin{equation*}
p\left(\bigcap\limits_{m \ne 1} {\left\{ {{X^n}(m) \ne {X^n}(1)} \right\}} \right) \le p\left(\left\{ {({U^n},{X^n}(1)) \notin A_\varepsilon ^{(n)}(p(u,x))} \right\} \cup \left\{ {\bigcap\limits_{m \ne 1} {\left\{ {{X^n}(m) \ne {X^n}(1)} \right\}} } \right\}\right).
\end{equation*}

\end{appendices}


\end{document}